\documentclass[letterpaper,twocolumn,10pt]{article}
 
\usepackage{usenix2019_v3}
\usepackage{macros}
\usepackage[labelfont=bf]{caption}
\usepackage{subcaption}
\usepackage[font={small,it}]{caption}
\usepackage{color}
\usepackage{tikz}
\usepackage{amsmath}
\usepackage{subcaption}
\usepackage{graphicx}
\usepackage{tabularx}
\usepackage{array,multirow,rotating}
\usepackage{algorithm}
\usepackage{algpseudocode}
\usepackage{amsmath}
\usepackage{amssymb}
\usepackage{tikz}

\usepackage{mathtools}

\usepackage[toc,page]{appendix} 

\usepackage{pifont}
\newcommand{\cmark}{\ding{51}}%
\newcommand{\xmark}{\ding{55}}%

\usepackage[compact]{titlesec}
\microtypecontext{spacing=nonfrench}
\titleformat{\subsection}
  {\normalfont\fontsize{10}{10}\bfseries}{\thesubsection}{1em}{}

\usepackage[inline]{enumitem}

\newcommand*\tcircled[1]{\tikz[baseline=(char.base)]{
            \node[shape=circle,draw,thick,font=\bf,inner sep=1pt] (char) 
            {\footnotesize#1};}}
            
\usepackage{wrapfig}
\usepackage{soul}

 \sloppypar
\def\footnoterule{\kern-3pt
  \hrule \kern 2.6pt} 
\usepackage{titling}
\date{}
\usepackage{lipsum}
\usepackage{booktabs}

\let\OLDthebibliography\thebibliography
\renewcommand\thebibliography[1]{
  \OLDthebibliography{#1}
  \setlength{\parskip}{0pt}
  \setlength{\itemsep}{1pt plus 0.2ex}
}
\usepackage{soul}
\usepackage{fancyhdr}
\begin{document}
\title{\Large \bf Cocktail: Leveraging Ensemble Learning for Optimized Model Serving\\ in Public Cloud.}
\author{
{\rm Jashwant Raj Gunasekaran}\\
\small Pennsylvania State University \and
{\rm Cyan Subhra Mishra}\\
\small Pennsylvania State University \and
{\rm Prashanth Thinakaran}\\
\small Pennsylvania State University \and
{\rm Mahmut Taylan Kandemir}\\
\small Pennsylvania State University \and
{\rm Chita R. Das}\\
\small Pennsylvania State University
}

\maketitle
 \sloppypar

\def\footnoterule{\kern-3pt
  \hrule \kern 2.6pt} 

\begin{abstract}
\vspace{-1mm}
With a growing demand for adopting ML models for a variety of application services, it is vital that the frameworks serving these models are capable of delivering highly accurate predictions with minimal latency along with reduced deployment costs in a public cloud environment. Despite high latency, prior works in this domain are crucially limited by the accuracy offered by individual models. Intuitively, model ensembling  can address the accuracy gap by intelligently combining different models in parallel. However, selecting the appropriate models dynamically at runtime to meet the desired accuracy with low latency at minimal deployment cost is a nontrivial problem.
Towards this, we propose \emph{Cocktail}, a cost effective ensembling-based model serving framework. {\em Cocktail} comprises of two key components: (i) a dynamic model selection framework, which reduces the number of models in the ensemble, while satisfying the accuracy and latency requirements; (ii) an adaptive resource management (RM) framework that employs a distributed proactive autoscaling policy combined with importance sampling, to efficiently allocate resources for the models. The RM framework leverages transient virtual machine (VM) instances  to reduce the deployment cost in a public cloud. A prototype implementation of \emph{Cocktail} on the AWS EC2 platform and exhaustive evaluations using a variety of workloads demonstrate that \emph{Cocktail} can reduce deployment cost by 1.45$\times$, while providing 2$\times$ reduction in latency and satisfying the target accuracy for up to 96\% of the requests, when compared to state-of-the-art model-serving frameworks. 


\end{abstract}

\section{Introduction} 
\label{sec:intro}


Machine Learning (ML) has revolutionized user experience in various cloud-based application domains such as product recommendations~\cite{shaya2010intelligent}, personalized advertisements~\cite{8327042}, and computer vision~\cite{bartlett2005recognizing,sirius}. For instance, Facebook~\cite{8327042,wu2019machine} serves trillions of inference requests for user-interactive applications like ranking new-feeds, classifying photos, etc. It is imperative for these applications to deliver accurate predictions at sub-millisecond latencies~\cite{swayam,clipper,infaas,clockwork,8327042,deeprecsys} as they critically impact the user experience. This trend is expected to perpetuate as a number of applications adopt a variety of ML models to augment their services. These ML models are typically trained and hosted on cloud platforms as service end-points, also known as \textit{model-serving} framework~\cite{sagemaker,serving,deepstudio}. From the myriad of ML flavours, Deep Neural Networks  (DNNs)~\cite{DNN} due to their multi-faceted nature, and highly generalized and accurate learning patterns~\cite{tan2019efficientnet,howard2019searching} are dominating the landscape by making these model-serving frameworks accessible to developers. However, their high variance due to the fluctuations in training data along with compute and memory intensiveness~\cite{NetAdapt,rahman2016efficient,narayanan2019pipedream} has been a major impediment in designing models with high accuracy and low latency. Prior model-serving frameworks like InFaas~\cite{infaas} are confined by the accuracy and latency offered by such individual models.

Unlike single-model inferences, more sophisticated techniques like {\em ensemble learning}~\cite{ensemble} have been instrumental in allowing model-serving to further improve accuracy with multiple models. For example, by using the ensembling~\footnote{We refer to ensemble-learning as ensembling throughout the paper.} technique, images can be classified using multiple models \emph{in parallel} and results can be combined to give a final prediction. This significantly boosts accuracy compared to single-models, and for this obvious advantage, frameworks like Clipper~\cite{clipper} leverage ensembling techniques. Nevertheless, with ensembling, the very high resource footprint due to sheer number of models that need to be run for each request 
~\cite{clipper,10.1145/1835804.1835914}, exacerbates the public cloud deployment costs, as well as leads to high variation in latencies. Since cost 
plays a crucial role in application-provider consideration, it is quintessential to minimize the deployment costs, while maximizing accuracy with low latency. Hence, the non-trivial challenge here lies in making the cost of ensembling predictions analogous to single model predictions, while satisfying these requirements.
Studying the state-of-the-art ensemble model-serving frameworks, we observe the following critical shortcomings:  

$\bullet$ 
Ensemble model selection policies used in frameworks like Clipper~\cite{clipper} are static, as they \emph{ensemble all available models} and focus solely on minimizing loss in accuracy. This leads to higher latencies and further inflates the resource footprint, thereby accentuating the deployment costs.

$\bullet$ Existing ensemble weight estimation~\cite{MAB} has \emph{high computational complexity} and, in practice, are limited to a small set of off-the-shelf models. This leads to significant loss in accuracy. Besides, employing linear ensembling techniques such as model averaging are compute intensive~\cite{rafiki} and not scalable for a large number of available models.

$\bullet$ Ensemble systems~\cite{rafiki,clipper} are \emph{not focused towards model deployment} in a public cloud infrastructure, where resource selection and procurement play a pivotal role in minimizing the latency and deployment costs. Further, the resource provisioning strategies employed in single model-serving systems are {\em not directly extendable} to ensemble systems.  

These shortcomings collectively motivate the central premise of this work: \emph{how to solve the complex optimization problem of cost, accuracy and latency for an ensembling framework?} In this paper, we present and evaluate \emph{Cocktail}\footnote{Cocktail is ascribed to having the perfect blend of models in an ensemble.}, which to our knowledge is the first work that develops a cost-effective model-serving system by exploiting ensembling techniques to deliver high accuracy and low latency predictions in public cloud. \emph{Cocktail} adopts a three-pronged approach to solve the optimization problem. First, it uses a dynamic model selection policy to significantly reduce the number of models used in an ensemble, while meeting the latency and accuracy requirements. \begin{wrapfigure}{l}{0.24\textwidth}
\centering
\includegraphics[width=0.24\textwidth]{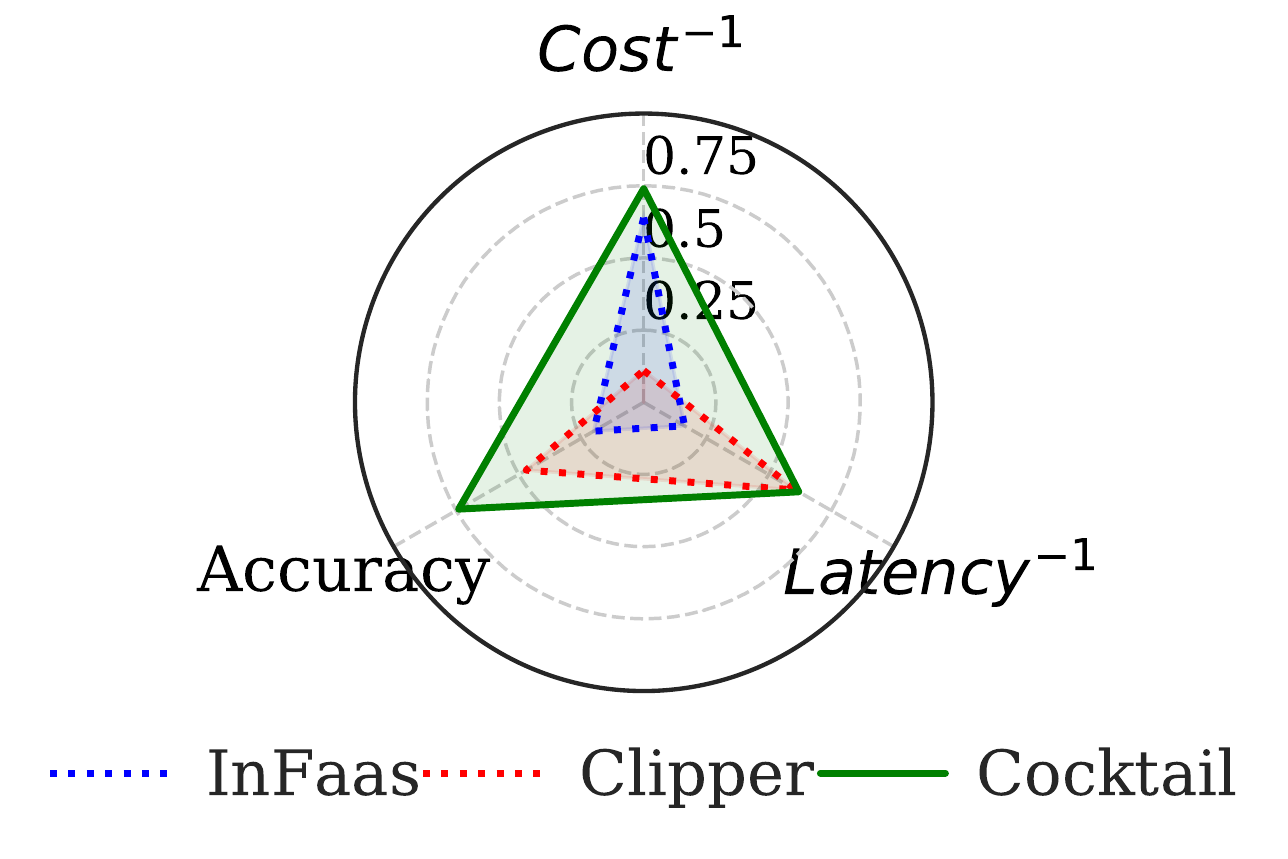}
\caption{Benefits of \emph{Cocktail}. Results are normalized (higher the better).}
\label{fig:3d}
\end{wrapfigure}Second, it utilizes distributed autoscaling policies to reduce the latency variability and resource consumption of hosting ensemble models. Third, it minimizes the cost of deploying ensembles in a public cloud by taking advantage of transient VMs, as they can be 70-90\% cheaper~\cite{ali2019spotweb} than traditional VMs. \emph{Cocktail}, by coalescing these benefits, is capable of operating in a region of optimal cost, accuracy and latency (shown in Figure~\ref{fig:3d}) that prior works cannot achieve.  
Towards this, the \textbf{key contributions} of the paper are summarized below:
\begin{enumerate}[leftmargin=*]
\itemsep-0.2em 
\item 
By characterizing accuracy \emph{vs.} latency of ensemble models, we identify that prudently selecting a subset of available models under a given latency can achieve the target accuracy. We leverage this in \emph{Cocktail}, to design a novel dynamic model selection policy, which ensures accuracy with significantly reduced number of models. 

\item Focusing on classification-based inferences, it is important to effectively minimize the bias in predictions resulting from multiple models. In \emph{Cocktail}, we employ a per-class weighted majority voting policy, that makes it scalable and effectively breaks ties when compared to traditional weighted averaging, thereby minimizing the accuracy loss. 
\item We show that uniformly scaling resources for all models in the ensemble leads to over-provisioning of resources and towards minimizing it, we build a distributed weighted auto-scaling policy that utilizes the \emph{importance sampling} technique to proactively allocate resources to every model. Further, \emph{Cocktail} leverages transient VMs as they are cheaper, to drastically minimize the cost for hosting model-serving infrastructure in a public cloud. 
 \item We implement a prototype of \emph{Cocktail} using both CPU and GPU instances on AWS EC2~\cite{EC2} platform and extensively evaluate it using different request-arrival traces. Our results from exhaustive experimental analysis demonstrate that \emph{Cocktail} can minimize deployment cost by 1.4$\times$ while meeting the accuracy for up-to 96\% of the requests and providing 2$\times$ reduction in latency, when compared to state-of-the-art model serving systems. 
\item We show that ensemble models are inherently fault-tolerant over single models, since in the former, failure of a model would incur some accuracy loss without complete failure of the requests. It is observed from our failure-resilience results that \emph{Cocktail} can adapt to instance failures by limiting the accuracy loss within 0.6\%.\vspace{-2mm}
\end{enumerate}
\begin{figure}
\centering
\begin{minipage}[t]{0.9\linewidth}
\centering
\includegraphics[width=0.8\textwidth]{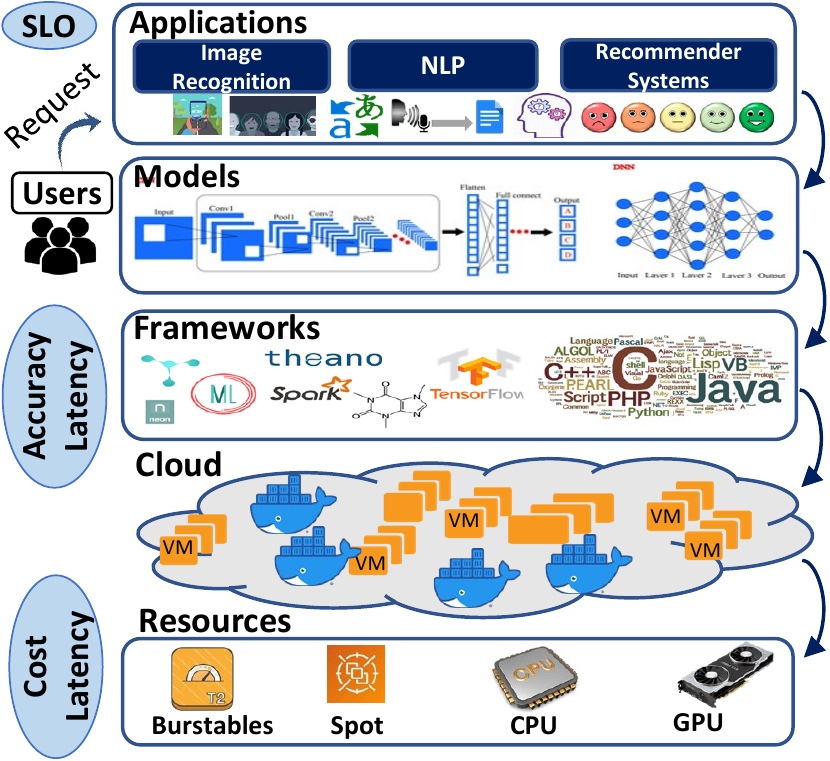}
\end{minipage}
\caption{The overall framework for model-serving in public cloud.}
\label{fig:model-serving}
\end{figure}


\section{Background and Motivation} 
\label{sec:background}

We start by providing a brief overview of model-serving in public cloud and ensembling, followed by a detailed analysis of their performance to motivate the need for \emph{Cocktail}. 

\subsection{Model Serving in Public Cloud} 
Figure~\ref{fig:model-serving} shows the overall architecture of a model-serving framework. There are diverse applications that are typically developed, trained and hosted as web services. These services allow end-users to submit queries via web server interface. {\color{black} Since these inference requests are  often user-facing, it is imperative to administer them under a strict service level objective (SLO). {We define SLO as the end-to-end response latency required by an application.} Services like Ads and News Feed~\cite{8327042,deeprecsys} would require SLOs within 100ms, while facial tag recommendation~\cite{infaas} can tolerate up to 1000ms.} A myriad of model architectures are available to train these applications which by themselves can be deployed on application frameworks like  \texttt{TensorFlow}~\cite{tensorflow}, \texttt{PyTorch}~\cite{pytorch} etc. Table~\ref{tbl:models} shows the different models available for image prediction, that are pretrained on Keras using \texttt{ImageNet}~\cite{imagenet} dataset. Each model has unique accuracy and latencies depending on the model architecture. Typically denser models are designed with more parameters (ex. \texttt{NASLarge}) to classify complex classes of images. 

The entire model framework is typically hosted on resources like VMs or containers in public cloud. These resources are available in different types including but not limited to transient instances, burstables, CPUs, and GPUs. Transient instances~\cite{spotcheck} are similar to traditional VMs but can be revoked at any time by the cloud provider with an interruption notice. 
The provisioning latency, instance permanence and packing factor of these resources have a direct impact on the latency and cost of hosting model-serving. We explain instance ``packing factor'' and its relationship with latency in Section~\ref{sec:ensemble-overhead}. In this paper, we focus on improving the accuracy and latency from the model selection perspective and consider instances types from a cost perspective.
\begin{table}
\centering
\footnotesize
\resizebox{0.45\textwidth}{!}{%
\begin{tabular}{||l|c|c|c|l||}
\hline
\textbf{Model (Acronym)} & \textbf{\begin{tabular}[c]{@{}c@{}}Params \\ (10k)\end{tabular}} & \textbf{\begin{tabular}[c]{@{}c@{}}Top-1\\ Accuracy(\%)\end{tabular}} & \textbf{\begin{tabular}[c]{@{}c@{}}Latency \\ (ms)\end{tabular}} & \textbf{\begin{tabular}[c]{@{}l@{}}$P_f$\end{tabular}} \\ \hline
MobileNetV1 (MNet) & 4,253 & 70.40 & 43.45 & 10 \\ \hline
MobileNetV2 (MNetV2) & {\color[HTML]{263238} 4,253} & {\color[HTML]{263238} 71.30} & 41.5 & 10 \\ \hline
NASNetMobile (NASMob) & 5,326 & {\color[HTML]{263238} 74.40} & 78.18 & 3 \\ \hline
DenseNet121 (DNet121) & {\color[HTML]{263238} 8,062} & {\color[HTML]{263238} 75.00} & 102.35 & 3 \\ \hline
DenseNet201 (DNet201)& {\color[HTML]{263238} 20,242} & {\color[HTML]{263238} 77.30} & 152.21 & 2 \\ \hline
Xception (Xcep)& 22,910 & {\color[HTML]{263238} 79.00} & 119.2 & 4 \\ \hline
Inception V3 (Incep) & {\color[HTML]{263238} 23,851} & {\color[HTML]{263238} 77.90} & 89 & 5 \\ \hline
ResNet50-V2 (RNet50) & {\color[HTML]{263238} 25,613} & {\color[HTML]{263238} 76.00} & 89.5 & 6 \\ \hline
Resnet50 (RNet50) & {\color[HTML]{263238} 25,636} & {\color[HTML]{263238} 74.90} & 98.22 & 5 \\ \hline
IncepResnetV2 (IRV2)& 55,873 & {\color[HTML]{263238} 80.30} & 151.96 & 1 \\ \hline
NasNetLarge (NasLarge) & 343,000 & {\color[HTML]{263238} 82.00} & 311 & 1 \\ \hline
\end{tabular}
}
\caption{Collection of pretrained models used for image classification.} 
\label{tbl:models}
\end{table}
A majority of the model serving systems~\cite{mark,infaas,sagemaker} in public cloud support individual model selection from available models. For instance, InFaas~\cite{infaas} can choose variants among a same model to maintain accuracy and latency requirements. However, denser models tend to have up to 6$\times$ the size and twice the latency of smaller models to achieve increased accuracy of about 2-3\%. Besides using dense models, ensembling~\cite{ensemble} techniques have been used to achieve higher accuracy. \\
\textbf{Why Ensembling?} An Ensemble is defined as a set of classifiers whose individual decisions combined in some way to classify new examples. 
This has proved to be more accurate than traditional single large models because it inherently reduces incorrect predictions due to variance and bias. The commonly used ensemble method in classification problems are bootstrap aggregation (bagging) that considers homogeneous weak learners, learns them independently from each other in parallel, and combines them following some kind of deterministic averaging process~\cite{averaging} or majority voting~\cite{majority} process. For further details on ensemble models, we refer the reader to prior works~\cite{ensemble1,ensemble2,ensemble3,ensemble4,ensemble5,ensemble6,ensemble7}.
\subsection{Related Work}
{\color{black}\noindent{\textbf{Ensembling in practice}}: Ensembling is supported by commercial cloud providers like Azure ML-studio~\cite{azure-enemble} and AWS Autogluon~\cite{autogluon} to boost the accuracy compared to single models. Azure initially starts with 5 models and constantly scales up models using a hill-climb policy~\cite{caruana2004ensemble} to meet the target accuracy. While AWS combines all available (6-10) models to give the best possible accuracy. Users also have the option to manually mention the ensemble size. Unlike them, \emph{Cocktail's} model selection policy tries to right-size the ensemble for a given latency, while maximizing accuracy.} \\
\textbf{Model-serving in Cloud}: The most relevant prior works to \emph{Cocktail} are InFaas~\cite{infaas} and Clipper~\cite{clipper}, which have been extensively discussed and compared to in Section~\ref{sec:results}. 
{\color{black} Recently FrugalML~\cite{Chen2020FrugalML} was proposed to cost-effectively choose from commercial MLaaS APIs. While striking a few similarities with \emph{Cocktail}, it is practically limited to image-classification applications with very few classes and does not address resource provisioning challenges.} MArk~\cite{mark} proposed SLO and cost aware resource procurement policies for model-serving. Although our heterogeneous instance procurement policy has some similarities with MArk, it is significantly different because we consider ensemble models. 
Rafiki~\cite{rafiki} considers small model sets and scales up and down the ensemble size by trading off accuracy to match throughput demands. However, \emph{Cocktail's} resource management is more adaptive to changing request loads and does not drop accuracy. Pretzel~\cite{pretzel} and Inferline~\cite{inferline} are built on top of Clipper to optimize the prediction pipeline and cost due to load variations, respectively. Many prior  works~\cite{velox,laser,clockwork,swift} have extensively tried to reduce model latency by reducing overheads due to shared resources and hardware interference. We believe that our proposed policies can be complementary and beneficial to these prior works to reduce the cost and resource footprint of ensembling. There are mainstream commercial systems which automate single model-serving like TF-Serving~\cite{serving}, SageMaker~\cite{sagemaker}, AzureML~\cite{azureML}, Deep-Studio~\cite{deepstudio} etc.
\begin{table}
\footnotesize
\begin{center}
\resizebox{0.45\textwidth}{!}{%
 \begin{tabular}{||c | c | c | c |c | c | c |c||} 
 \hline
 \textbf{Features} & \begin{turn}{90}Clipper \cite{clipper}\end{turn} &
  \begin{turn}{90}Rafiki\cite{rafiki}\end{turn} &
  \begin{turn}{90}Infaas \cite{infaas}\end{turn} & \begin{turn}{90}MArk \cite{mark}\end{turn} &
 \begin{turn}{90}Sagemaker\end{turn} &
 \begin{turn}{90}Swayam \cite{swayam}\end{turn}& \begin{turn}{90}\emph{Cocktail} \end{turn} \\
 \hline
 \textbf{Predictive Scaling} & \xmark & \xmark &  \xmark & \cmark & \xmark & \cmark & \cmark \\  
 \hline
  \textbf{SLO Guarantees} & \cmark & \xmark &  \cmark & \cmark & \xmark & \cmark & \cmark \\
 \hline
  \textbf{Cost Effective} & \xmark & \xmark & \cmark & \cmark & \xmark & \xmark & \cmark \\
   \hline
   \textbf{Ensembling} & \cmark & \cmark & \xmark & \xmark &\cmark & \xmark & \cmark \\
   \hline
 \textbf{Heterogeneous Instances} & \xmark & \cmark &  \cmark & \cmark & \cmark &\xmark & \cmark \\
  \hline
 \textbf{Dynamic ensemble selection} & \xmark & \xmark & \xmark & \xmark &\xmark & \xmark & \cmark \\
   \hline
\textbf{Model abstraction} & \cmark & \cmark & \cmark & \xmark &\xmark & \xmark & \cmark \\
\hline
\end{tabular}}
\end{center}
\caption{{\color{black}Comparing {\emph{Cocktail}} with other related frameworks.}}
  \label{tbl:related}
\end{table}
\\
\textbf{Autoscaling in Public Cloud}: There are several research works that optimize the resource provisioning cost in public cloud. These works are broadly categorized into: (i) multiplexing the different instance types (e.g., Spot,  On-Demand)~\cite{swayam,tributary,proteus,stratus,Wang:2017:UBI:3107080.3084448,burscale,exosphere}, (ii) proactive resource provisioning based on prediction policies~\cite{swayam,mark,10.5555/2748143.2748357,spotcheck,tributary}. \emph{Cocktail} uses similar load prediction models and auto-scales VMs in a distributed fashion with respect to model ensembling. Swayam~\cite{swayam} is relatively similar to our work as it handles container provisioning and load-balancing, specifically catered for single model inferences. \emph{Cocktail's} autoscaling policy strikes parallels with Swayam's distributed autoscaling;  however, we further incorporate novel importance sampling techniques to reduce over-provisioning for under-used models. Table~\ref{tbl:related} provides a comprehensive comparison of \emph{Cocktail} with the most relevant works across key dimensions.  
\subsection{Pros and Cons of Model Ensembling}
In this section, we quantitatively evaluate (i) how effective ensembles are in terms of accuracy and latency compared to single models, and (ii) the challenges in deploying ensemble frameworks in a cost-effective fashion on a public cloud. For relevance in comparison to prior work~\cite{clipper,infaas} we chose image inference as our ensemble workload. While ensembling is applicable in other classification tasks like product recommendations~\cite{recommender,videorecommend}, text classification~\cite{nlp} etc, the observations drawn are generic and applicable to other applications.

\begin{table}
\centering
\scriptsize
\begin{tabular}{|p{1.2cm}|p{1cm}|p{0.85cm}|p{0.9cm}|p{0.85cm}|p{0.9cm}|}
\hline

\textbf{Baseline(BL)} & \textbf{NASLarge} & \textbf{IRV2} & \textbf{Xception} & \textbf{DNet121} & \textbf{NASMob}\\ \hline
\textbf{\#Models}& {10} & {8} & {7} & {5} & {2} \\ \hline
\textbf{BL\_Latency}& {311(ms)} & {152(ms)} & {120(ms)} & {100(ms)} & {98(ms)} \\ \hline
\textbf{E\_Latency}&{152(ms)} & {120(ms)} & {103(ms)} & {89(ms)} & {44(ms)} \\ \hline
\end{tabular}
\caption{{\color{black} Comparing latency of Ensembling (E\_Latency) with single (baseline) models.}}
\label{tbl:ensemble}
\end{table}
\subsubsection{Ensembling Compared to Single Models}
To analyze the accuracy offered by ensemble models, we conduct an experiment using 10000 images from \texttt{ImageNet}~\cite{imagenet} test dataset, on a \texttt{C5.xlarge}~\cite{c5} instances in AWS EC2~\cite{EC2}. {\color{black} For a given baseline model, we combine all models whose latency is lower than that of the 
baseline, and call it full-ensemble. We perform ensembling on the predictions using a simple majority voting policy. The latency numbers for the baseline models and the corresponding ensemble models along with the size of the ensemble are shown in Table~\ref{tbl:ensemble}}. 
In majority voting, every model votes for a prediction for each input, and the final output prediction is the one that receives more than half of the votes. Figure~\ref{fig:top5}, shows the accuracy comparison of the baseline (single) and static ensemble (explained in Section~\ref{sec:static}) compared to the full-ensemble. {\color {black} It is evident that full-ensemble can achieve up to 1.65\% better accuracy than single models. }

Besides accuracy again, ensembling can also achieve lower latency. The latency of the ensemble is calculated as the time between start and end of the longest running model.{\color{black}  As shown in Table~\ref{tbl:ensemble}, in the case of \texttt{NASLarge}, the ensemble latency is 2$\times$ lower (151ms) than the baseline latency (311ms)}. {\color{black} Even a 10ms reduction in latency is of significant importance to the providers~\cite{clockwork}.}  {\color{black}  We observe a similar trend of higher ensemble accuracy for other four baseline models with a latency reduction of up to 1.3$\times$. }Thus, depending on the model subset used in the ensemble, it achieves better accuracy than the baseline at lower latencies. Note that in our example model-set, the benefits of ensembling will diminish for lower accuracies (< 75\%) because single models can reach those accuracies. Hence, based on the user constraints, \emph{Cocktail} chooses between ensemble and single models.


\subsubsection{Ensembling Overhead}
\label{sec:ensemble-overhead}
While ensembling can boost accuracy with low latency, their distinctive resource hungry nature drastically increases the deployment costs when compared to single models. This is because more VMs or containers have to be procured to match the resource demands. However, note that the ``Packing factor'' ($P_f$) for each model also impacts  the deployment costs. $P_f$ in this context is defined as the number of inferences that can be executed concurrently in a single instance without violating the inference latency (on average). Table~\ref{tbl:models} provides the $P_f$ for 11 different models when executed on a \texttt{C5.xlarge} instance. There is a linear relationship between $P_f$ and the instance size. It can be seen that smaller models \texttt{(MNet, NASMob}) can be packed 2-5$\times$ more when compared to larger models \texttt{(IRV2, NASLarge)}. Thus, the ensembles with models of higher $P_f$ have significantly lower cost.

\begin{figure}
\centering
\begin{minipage}[t]{0.9\linewidth}
\begin{subfigure}[t]{0.49\textwidth}
\centering
 \includegraphics[width=0.9\textwidth]{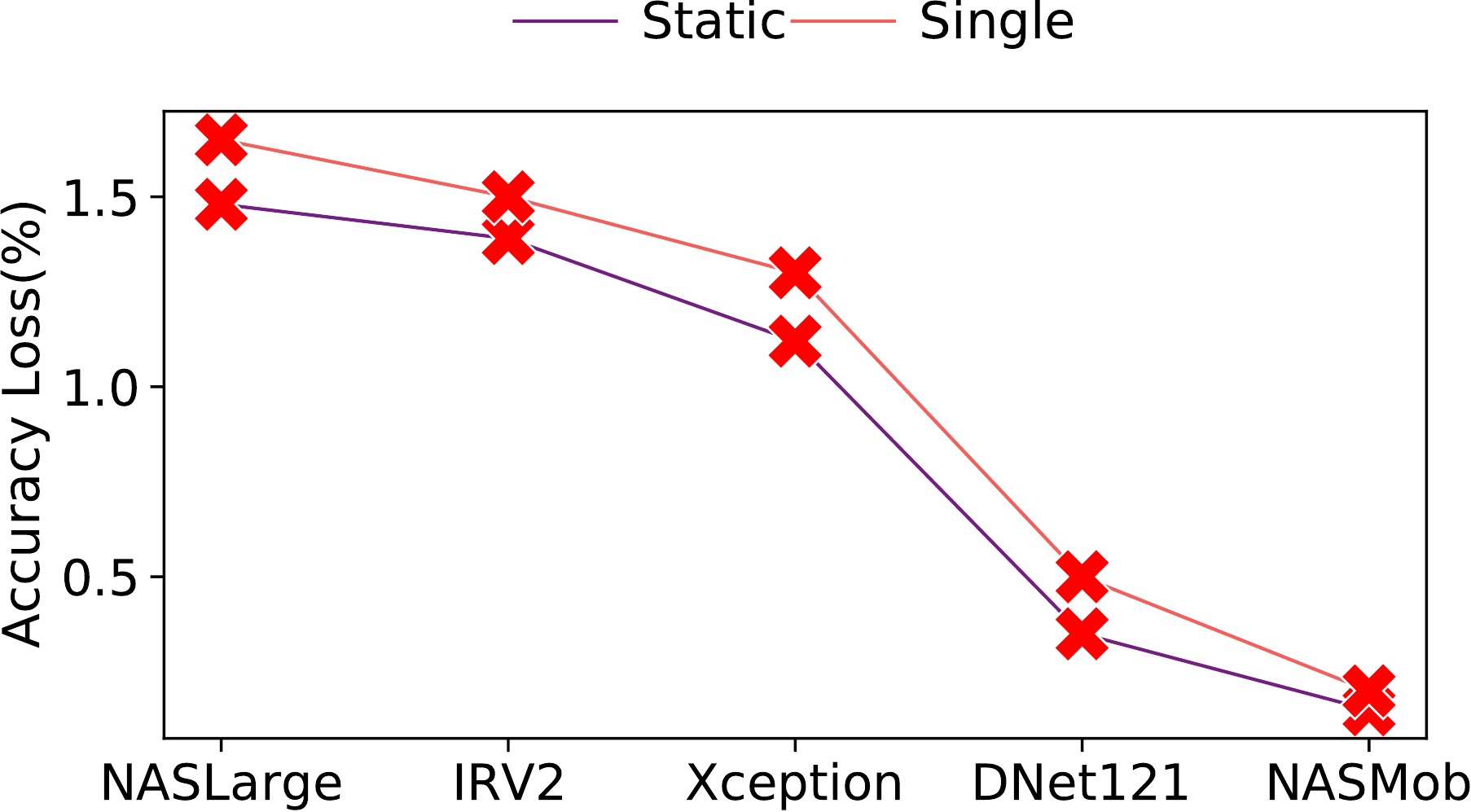}
\caption{Accuracy loss compared to full-ensemble.}
\label{fig:top5}
\end{subfigure}
\begin{subfigure}[t]{.49\textwidth}
\centering
\includegraphics[width=0.99\textwidth]{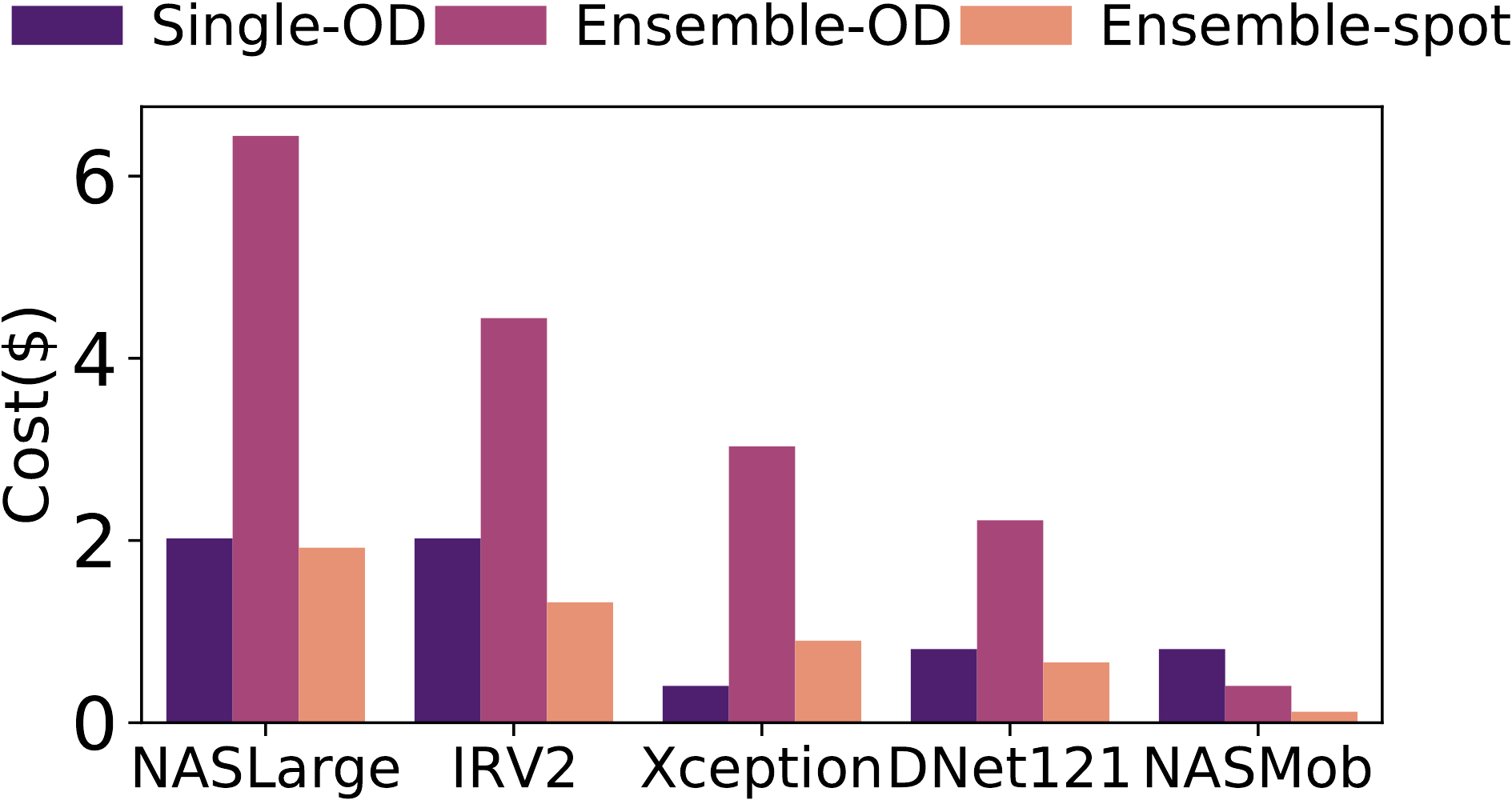}
\caption{Cost of full-ensembling hosted on OD and Spot instances.}
\label{fig:spot-cost}
\end{subfigure}
\end{minipage}
\caption{Comparing the cost and accuracy of ensembling to single models.}
\label{fig:motivation}
\end{figure}

The benefits of $P_f$ is contingent upon the models chosen by the model selection policy. Existing ensemble model selection policies used in systems like Clipper use all off-the-shelf models and assign weights to them to calculate accuracy. However, they do not right-size the model selection to include models which primarily  contribute to the majority voting. {\color{black} We compare the cost of hosting ensembles using both spot (ensemble-spot) and OD (ensemble-OD) instances with the single models hosted on OD (single-OD) instances. Ensemble-spot is explained further in the next section. We run the experiment over a period of 1 hour for 10 requests/second. The cost is calculated as the cost per hour of EC2 c5.xlarge instance use, billed by AWS~\cite{EC2}. We ensure all instances are fully utilized by packing multiple requests in accordance to the $P_f$. }
As shown in Figure~\ref{fig:spot-cost}, Ensemble-OD is always expensive than single-OD for the all the models. 
Therefore, it is important to ensemble an ``optimal'' number of less compute intensive models to reduce the cost.
\section{Prelude to Cocktail}
\label{sec:modeling}

To specifically address the cost of hosting an ensembling-based model-serving framework in public clouds without sacrificing the accuracy, this section introduces an overview of the two primary design choices employed in \emph{Cocktail}.
\\
\textbf{How to reduce resource footprint?}
\label{sec:static}
The first step towards making model ensembling cost effective is to minimize the number of models by pruning the ensemble, which reduces the overall resource footprint. In order to estimate the right number of models to participate in a given ensemble, we conduct an experiment where we chose top $\frac{N}{2}$ accurate models (static) from the full-ensemble of size $N$. From Figure~\ref{fig:top5}, it can be seen that the static policy has an accuracy loss of up to 1.45\% when compared to full-ensemble, but is still better than single models. This implies that the models other than top $\frac{N}{2}$ yields a significant 1.45\% accuracy improvement in the full-ensemble but they cannot be statically determined. \begin{wrapfigure}{l}{0.22\textwidth}
\centering
\includegraphics[width=0.22\textwidth]{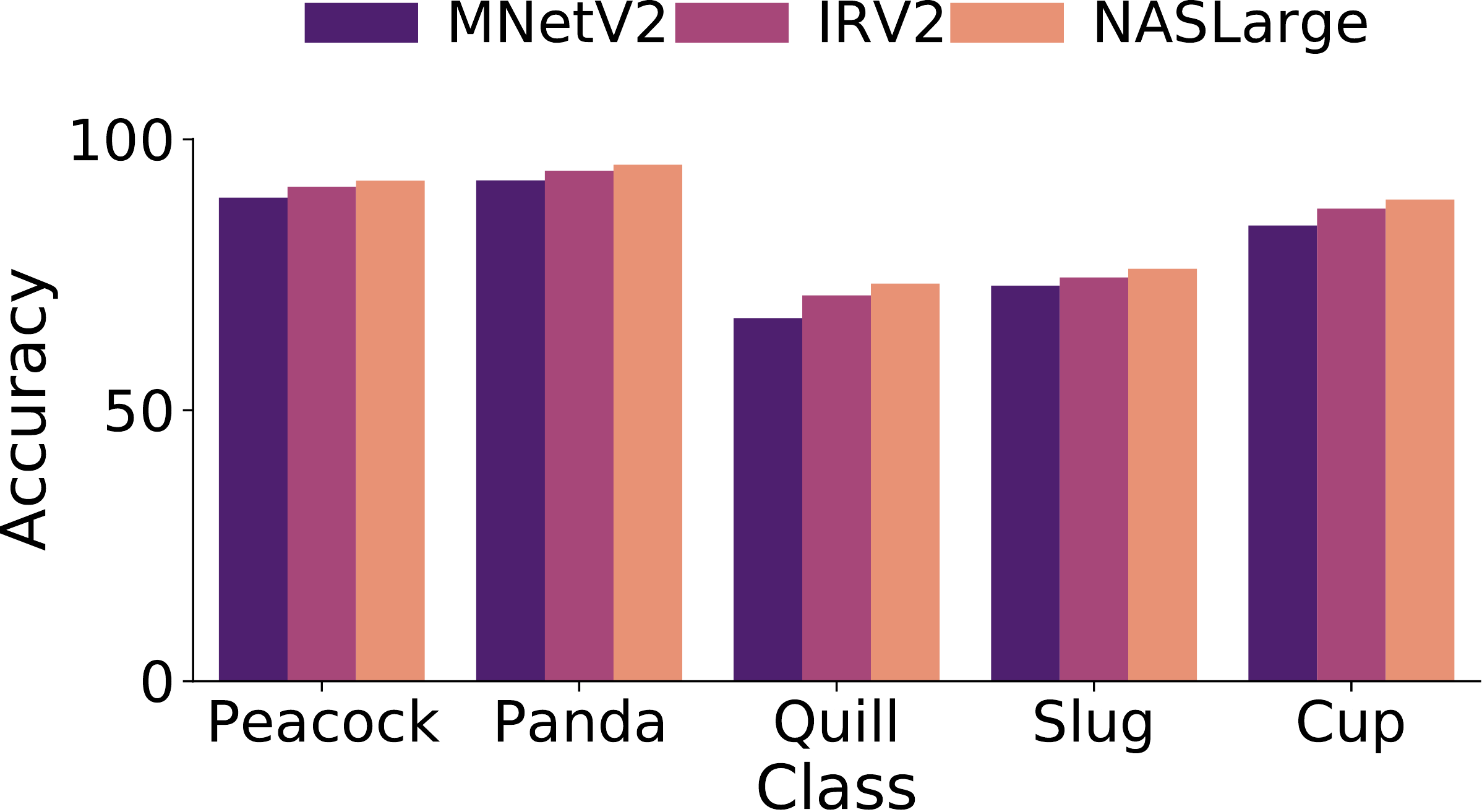}
\caption{{\color{black}Class-wise Accuracy.}}
\label{fig:class-wise}
\end{wrapfigure} Therefore, a full-ensemble model participation is not required for all the inputs because, every model is individually suited to classify certain classes of images when compared to other classes. {\color{black} Figure~\ref{fig:class-wise} shows the class-wise accuracy for three models on 5 distinct classes.  It can be seen that for simpler classes like Slug, \texttt{MNetV2} can achieve similar accuracy as the bigger models, while for difficult classes, like Cup and Quill, it experiences up to 3\% loss in accuracy.} Since the model participation for ensembling can vary based on the class of input images being classified, there is a scope to develop a dynamic model selection policy that can leverage this class-wise variability to intelligently determine the number of models required for a given input.\\
\textbf{Key Takeaway:} \emph{Full ensemble model-selection is an overkill, while static-ensemble leads to accuracy loss. This calls for a dynamic model selection policy which can accurately determine the number of models required, contingent upon the accuracy and scalability of the model selection policy.}\\
\textbf{How to save cost?} \label{sec:failure} Although dynamic model selection policies can significantly reduce the resource footprint as shown in Figure~\ref{fig:spot-cost}, the cost is still 20-30\% higher when compared to a single model inference. Most cloud providers offer transient VMs  such as Amazon Spot instances~\cite{spotcheck}, Google Preemptible VMs~\cite{google-preempt}, and Azure Low-priority VMs~\cite{azure-batch}, that can reduce cloud computing costs by as much as 10$\times$~\cite{ali2019spotweb}. In \emph{Cocktail}, we leverage these transient VMs such as spot instances to drastically reduce the cost of deploying ensembling model framework. {\color{black} As an example, we host full-ensembling on AWS spot instances. Figure~\ref{fig:spot-cost}  shows that ensemble-spot can reduce the cost by up to 3.3$\times$ when compared to ensemble-OD. For certain baselines like IRV2, ensemble-spot is also 1.5$\times$ cheaper than single-OD. }However, the crucial downside of using transient VMs is that they can be unilaterally preempted by the cloud provider at any given point due to reasons like increase in bid-price or provider-induced random interruptions. As we will discuss further, \emph{Cocktail} is resilient to instance failures owing to the fault-tolerance of ensembling by computing multiple inferences for a single request. \\ 
\textbf{Key takeaway}: \emph{The cost-effectiveness of transient instances, is naturally suitable for hosting ensemble models.}
\section{Overall Design of Cocktail}
\label{sec:scheme}

\begin{figure}[t]
\begin{minipage}{0.99\linewidth}
\centering
\includegraphics[width=0.99\textwidth]{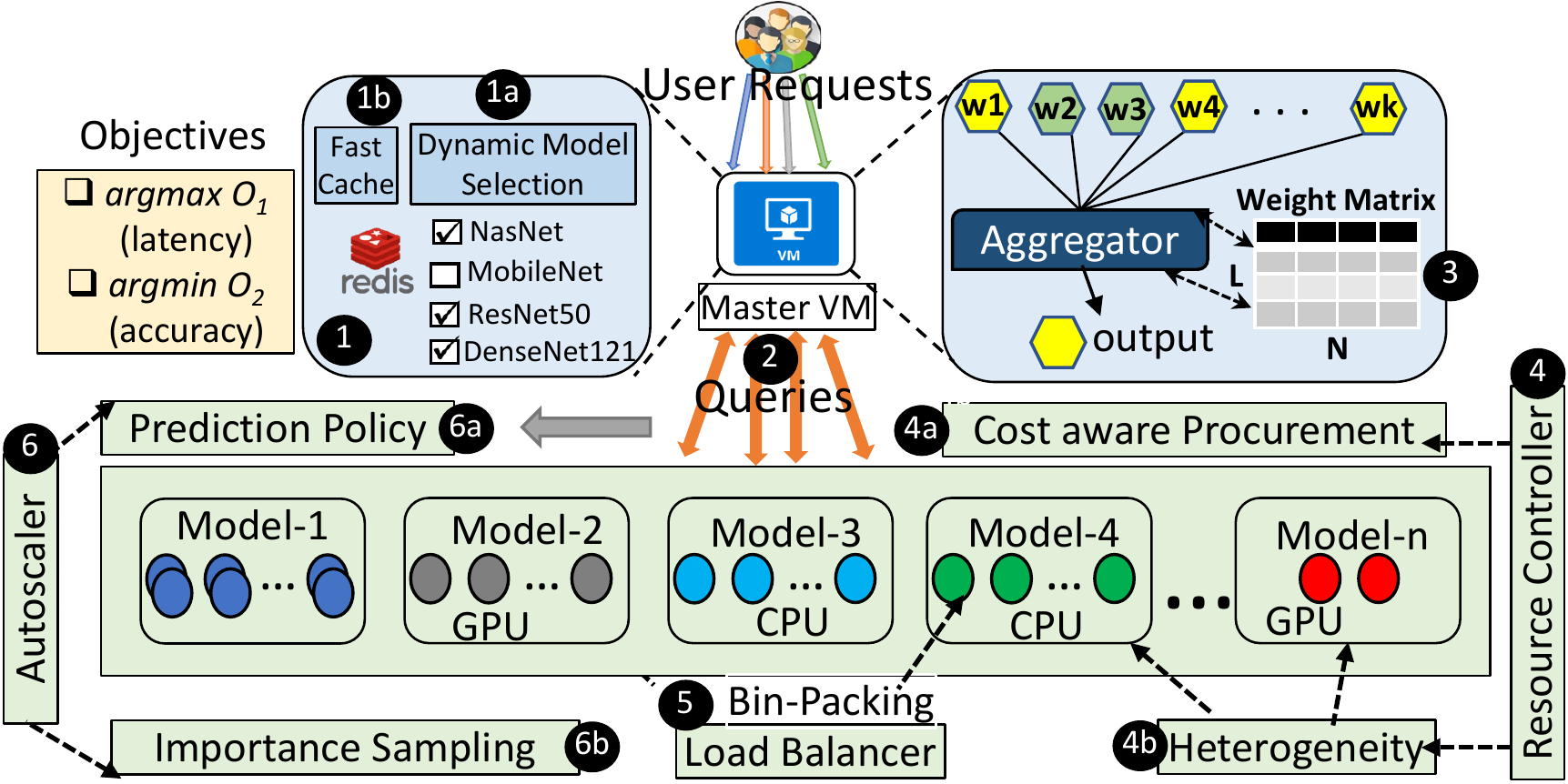}
\end{minipage}
\caption{High-level overview of \emph{Cocktail} design.}
\label{fig:Cocktail}
\end{figure}
Motivated by our observations, we design a novel model-serving framework, \emph{Cocktail}, that can deliver high-accuracy and low-latency predictions at reduced cost. Figure~\ref{fig:Cocktail} depicts the high-level design of \emph{Cocktail}. Users submit requests to a master VM, which runs a model selection algorithm, ~\circled{1a} to decide the models to participate in the ensemble. The participating models are made available in a model cache~\circled{1b} for faster access and avoid re-computation for requests having similar constraints. Then, individual queries are dispatched to instances pools~\circled{2} dedicated for each model. The results from the workers are {\em ensembled} using an weighted majority voting aggregator~\circled{3} to agree upon a correct prediction. 
To efficiently address the resource management and scalability challenges, \emph{Cocktail} applies multiple strategies. First, it maintains dedicated instance pools to serve individual models which simplifies the management and load balancing overheads for every model. 
Next, the resource controller~\circled{4} handles instance procurement, by exploiting both CPU and GPU  instances~\circled{4a} in a cost-aware~\circled{4b} fashion, while the load balancer \circled{5} ensures all procured instances are bin-packed by assigning queries to appropriate instances. We also design an  autoscaler~\circled{6},  which utilizes a prediction policy~\circled{6a} to forecast the request load and scale instances for every model pool, thereby minimizing over-provisioning of resources. The autoscaler further employs an importance sampling~\circled{6b} algorithm to estimate the importance of each model pool by calculating percentage of request served by it in a given time interval. 
The key components of the design are explained in detail below. 
\subsection{Dynamic Model Selection Policy}
\label{sub:scheme:DynModel}
We use a window-based dynamic model selection policy using two objective functions as described below.\\
\textbf{Objective functions}: In order to reduce cost and latency while maximizing the accuracy, we define a latency-accuracy metric ($\mu_{AL}$) and cost metric ($\mu_{c}$):
\begin{align*}
\mu_{AL} &= \frac{Acc_{target}}{Lat_{target}}  & \mu_C &= k \times \sum_{m=1}^{N}\frac{inst\_cost}{P_{f_{m}}}
  \end{align*}
where $N$ is the number of models used to ensemble and ${inst\_cost}$ is the VM cost. Each model $m$ has a packing factor $P_{f_{m}}$ and k is a constant which depends on the VM size in terms of vCPUs (xlarge, 2xlarge, etc). {\color{black} Our first objective function ($O_1$) is to the maximize $\mu_{AL}$ such that target accuracy ($ Acc_{target}$) is reached within the target latency  ($Lat_{target}$).
\begin{align*}
\max \mu_{AL} : \left\{\begin{matrix}
 &Acc_{target} \ge Acc_{target} \pm Acc_{margin} \\ 
 & Lat_{target} \le Lat_{target} \pm Lat_{margin} 
\end{matrix}\right.
\end{align*}
To solve $O_1$, we determine an initial model list by choosing the individual models satisfying $Lat_{target}$ and then create a probabilistic ensemble that satisfies the $Acc_{target}$. \emph{Cocktail} takes the accuracy of each model as a probability of correctness and then iteratively constructs a model list, where the joint probability of them performing the classification is within the accuracy target.  We tolerate a 0.2\% ($Acc_{margin}$) and 5ms ($Lat_{margin}$) variance in $Acc_{target}$ and $Lat_{target}$, respectively.} Next, we solve for the second objective function ($O_2$) by minimizing $\mu_{C}$, while maintaining the  target accuracy.
\begin{align*}
\min
\mu_{C} : \left\{\begin{matrix}
 &Acc_{target} \ge Acc_{target} \pm Acc_{margin}
\end{matrix}\right.
\end{align*}
$O_2$ is solved by resizing the model list of size \emph{N} and further through intelligence resource procurement (described in section~\ref{sec:resource}), and thus maximizing $P_f$ and minimizing $k$ simultaneously. If we have $N$ models, where each model has a minimum accuracy of `$a$', we model this ensemble as a coin-toss problem, where $N$ biased coins (with probability of head being $a$) are tossed together, and we need to find the probability of majority of them being heads. For this, we need at least $\lfloor \frac{N}{2} \rfloor +1$ models to give the same results. The probability of correct prediction is given by 
\begin{align*}
\sum_{i=\lfloor \frac{N}{2}\rfloor + 1}^N {\binom{N}{i}} \hspace{1mm }a^{i} \hspace{1mm}(1-a)^{(N-i)}
\vspace{-2mm}
\end{align*}
\begin{algorithm}[t]
\footnotesize
\caption{Model Selection and Weighted Majority Voting}
\begin{algorithmic}[1]
\Procedure{Full\_Ensemble(ModelList, SLO)}{}
\For{model $\in$ ModelList}
\If{model.latency $\le$ SLO.latency}
\State Model.add(model)
\EndIf
\EndFor   \hspace{2mm} \tcircled{$O_1$}
\EndProcedure
\hrulefill
\Procedure{Dynamic\_Model\_Scaling(\textit{Models})}{}
\If{\textit{curr\_accuracy $\ge$ accuracy\_threshold}} 
\If{\textit{$max_{vote}$ > $\frac{N}{2}$ + 1}} \tcircled{$O_2$}
\State$to\_be\_dropped \gets max_{vote} - \frac{N}{2}+1$
\State$Models.drop(to\_be\_dropped)$
\EndIf
\Else
\State$addModel \gets find\_models(remaining\_models)$
\State$Models.append(addModel)$
\EndIf
\EndProcedure 
\hrulefill
\Procedure{Weighted\_Voting(\textit{Models})}{}
\For{\textit{model in $\forall Models$}}
\State$class \gets model.predicted\_class$
\State$weighted\_vote[class] += weights[model.class]$
\EndFor
\State $P_{class} \gets \max(weighted\_vote, key=class)$
\State$ return P_{class}$
\EndProcedure
\end{algorithmic}
\label{algo1}
\end{algorithm}
\textbf{Model Selection Algorithm:} To minimize $\mu_C$, we design a policy to downscale the number of models, if more than N/2+1 models vote for the same classification result. Algorithm~\ref{algo1} describes the overall design of the model selection policy~\circled{1a}. For every monitoring interval, we keep track of the accuracy obtained from predicting all input images within the interval. If the  accuracy of the interval reaches the threshold accuracy  (target + error\_margin), we scale down the number of available models in the ensemble. For consecutive sampling intervals, we calculate the \texttt{Mode} (most frequently occurring) of the majority vote received for every input. If the \texttt{Mode} is greater than needed votes $\lfloor N/2\rfloor+1$ we prune the models to $\lfloor N/2\rfloor+1$. While down-scaling, we drop the models with the least prediction accuracy in that interval. If there is a tie, we drop the model with least packing factor ($P_f$). It can so happen that dropping models can lead to drop in accuracy for certain intervals, because the class of images being predicted are different. In such cases, we up-size the models (one at a time) by adding most accurate model from the remaining unused models. 
\subsubsection{Class-based Weighted Majority Voting}
\label{sec:voting}
The model selection policy described above ensures that we only use the necessary models in the majority voting. In order to increase the accuracy of majority voting,  we design a weighted majority voting policy~\circled{3}. The weight matrix is designed by considering the accuracy of each model for each class, giving us a weight matrix of $L\times N$ dimension, where $L$ is the number of unique labels and $N$ is the number of models used in the ensemble. 
The majority vote is calculated as a sum of model-weights for each unique class in the individual prediction of the ensemble. For instance, if there are 3 unique classes predicted by all the ensemble models, we sum the weights for all models of the same class. The class with the  maximum weight ($P_{class}$) is the output of the majority vote. Hence, classes that did not get the highest votes can still be the final output if the models associated with that class has a higher weight, than the combined weights of highest voted class. 
{\color {black} Unlike commonly used voting policies which assign weights based on overall correct predictions, our policy incorporates class-wise information to the weights, thus making it more adaptable to different images classes. }

In order to determine the weight of every class, we use a per-class dictionary that keeps track of the correct predictions of every model per class. We populate the dictionary at runtime to avoid any inherent bias that could result from varying images over time. Similarly, our model selection policy is also changed at runtime based on correct predictions seen during every interval. 
An important concern in majority voting is tie-breaking. Ties occur when two sets of equal number of models predict a different result. The effectiveness of weighted voting in breaking ties is discussed in Section~\ref{sec:results}.

\subsection{Resource Management}
\label{sec:resource}
Besides model selection, it is crucial to design an optimized resource provisioning and management scheme to host the models cost-effectively. We explain in detail the resource procurement and autoscaling policy employed in \emph{Cocktail}.
\subsubsection{Resource Controller}
Resource controller determines the cost-effective  combination of instances to be procured. We explain the details below. \\
\textbf{Resource Types}: We use both CPU and GPU instances~\circled{4a} depending on the request arrival load. GPU instances are cost-effective when packed with a large batch of requests for execution. Hence, inspired from prior work~\cite{mark,clipper}, we design an adaptive packing policy such that it takes into account the number of requests to schedule at time $T$ and $P_{f}$ for every instance. The requests are sent to GPU instances only if the load matches the $P_{f}$ of the instance.\\ 
\begin{algorithm}[!t]
\footnotesize
\caption{Predictive Weighted Instance Auto Scaling}
\begin{algorithmic}[1]
\Procedure{Weighted\_Autoscaling(\textit{Stages})}{}

\State$\textit{Predicted\_load} \gets \textit{\textbf{DeepARN\_Predict}(\textit{load})}$
\For{\textit{every Interval}}
\For{\textit{model in $\forall Models$}}
\State$model_{weight} \gets get\_popularity(model)$
\State$Weight.append(model_{weight})$
\EndFor
\EndFor
\If{\textit{Predicted\_load $\ge$ Current\_load}}
\For{\textit{model in $\forall Models$}}

\State$\textit{I\_n} \gets \textit{(Predicted\_load - Current\_load)}\times model_{weight} $
 \State$\textit{launch\_workers}(\textit{est\_VMs})$
\State$\textit{model.workers.append}(\textit{est\_VMs})$
\EndFor
\EndIf
\EndProcedure 
\end{algorithmic}
\label{algo2}
\end{algorithm}
\textbf{Cost-aware Procurement}:
The cost of executing in a fully packed instance determines how expensive is each instance. Prior to scaling-up instances, we need to estimate the cost~\circled{4b} of running them along with existing instances. At any given time $T$, based on the predicted load ($L_p$) and running instances $R_N$, we use a cost-aware  greedy policy to determine the number of additional instances required to serve as $A_n = L_p - C_r$, where $C_r = \sum_{i=1}^{N}{P_{f_i}}$,
is the request load which can be handled with $R_N$.
To procure $A_n$ instances, we greedily calculate the least cost instance as $\min_{\forall i \in instances} {Cost_i \times A_n/P_{f_i} }$. Depending on the cost-effectiveness ratio of $A_n/P_{f_i}$, GPUs will be preferred over CPU instances. \\
\textbf{Load Balancer}:
Apart from procuring instances, it is quintessential to design a load balancing and bin-packing~\circled{5} strategy to fully utilize all the provisioned instances. We maintain a request queue at every model pool. In order to increase the utilization of all instances in a pool at any given time, the load balancer submits every request from the queue to the lease remaining free slots (viz. instance packing factor $P_f$). This is similar to an online bin-packing algorithm. We use an idle-timeout limit for 10 minutes to recycle unused instances from every model pool. Hence, greedily assigning requests enables early scale down of lightly loaded instances. 
\subsubsection{Autoscaler}
\label{sec:autoscaler}
Along with resource procurement, we need to autoscale instances to satisfy the incoming query load. Though reactive policies (used in Clipper and InFaas) can be employed which take into account metrics like CPU utilization~\cite{infaas}, these policies are slow to react when there is dynamism in request rates. Proactive policies with request prediction are know to have superior performance~\cite{mark} and can co-exist with reactive policies. In \emph{Cocktail}, we use a load prediction model that can accurately forecast the anticipated load for a given time interval. Using the predicted load~\circled{6a}, \emph{Cocktail} spawns additional instances, if necessary, for every instance pool. In addition, we sample SLO violations for every 10s interval and reactively spawn additional instances to every pool based on aggregate resource utilization of all instances. This captures SLO violations due to mis-predictions. \\
\textbf{Prediction Policy}:
\label{sec:prediction}
To effectively capture the different load arrival patterns, we design a DeepAR-estimator (DeepARest) based prediction model. 
{\color{black} We zeroed in on the choice of using DeepARest by conducting an in-depth comparison (Table~\ref{tbl:prediction}) of the accuracy loss when compared with other state-of-the-art traditional and ML-based prediction models used in prior works~\cite{prediction-models,mark}.} {\color {black} As shown in Algorithm~\ref{algo2}, for every model under a periodic scheduling interval of 1 minute ($T_s$), we use the \emph{Predicted}\_load ($L_p$) at time $T$ + $T_p$ and compare it with the \emph{current\_load} to determine the number of instances ($I_n$).}
\begin{wraptable}{r}{0.19\textwidth}
\footnotesize
\centering
\begin{tabular}{|l|l|}
\hline
\textbf{Model} & \textbf{RMSE} \\ \hline
MWA & 77.5 \\ \hline
EWMA & 88.25 \\ \hline
Linear R. & 87.5 \\ \hline
Logsitic R. & 78.34 \\ \hline
Simple FF. & 45.45 \\ \hline
DeepArEst & 26.67 \\ \hline
LSTM & 28.56 \\ \hline
\end{tabular}
\caption{{\color {black}Prediction models.}}
\label{tbl:prediction}
\end{wraptable}
$T_p$ is defined as the average launch time for new instances. ($T_s$) is set to 1 minute as it is the typical instance provisioning time for EC2 VMs. To calculate ($L_p$), we sample the arrival rate in adjacent windows of size $W$ over the past S seconds. Using the global arrival rate from all windows, the model predicts ($L_p$) for $T_p$ time units from $T$. $T_p$ is set to 10 minutes because it is sufficient time to capture the variations in long-term future. All these parameters are tunable based on the system needs.  \\ 
\textbf{Importance Sampling:} An important concern in autoscaling is that the model selection policy dynamically determines the models in the ensemble for a given request constraints. Autoscaling the instances equally for every model based on predicted load, would inherently lead to over-provisioned instances for under-used models. To address this concern, we design a weighted autoscaling policy which intelligently auto-scales instances for every pool based on the weights. {\color {black} As shown in Algorithm~\ref{algo2}, weights are determined by frequency in which a particular model is chosen for requests (\emph{get\_popularity}) with respect to other models in the ensemble. The weights are multiplied with the predicted load to scale instances \emph{(launch\_workers)} for every model pool.} We name this as an importance sampling~\circled{6b} technique, because the model pools are scaled proportional to their  popularity. 
\section{Implementation and Evaluation}
\label{sec:implementation}
We have implemented a prototype of \emph{Cocktail}
and deployed it on AWS EC2~\cite{EC2} platform 
for evaluating the design with . The details of the implementation are described below.
\subsection{{\emph{\emph{Cocktail}} Prototype Implementation}}
\emph{Cocktail} is implemented using 10KLOC of \texttt{Python}. We designed \emph{Cocktail} as a client-server architecture, where one master VM receives all the incoming requests which are sent to individual model worker VMs.\\ 
\textbf{Master-Worker Architecture}: The master node handles the major tasks such as (i) concord model selection policy, (ii) request dispatch to workers VMs as asynchronous future tasks using \texttt{Python} \texttt{asyncio} library, and (iii) ensembling the prediction from the worker VMs. Also all VM specific metrics such as current\_load, CPU utilization, etc. reside in the master node. It runs on a \texttt{ C5.16x}~\cite{c5} large instance to handle these large volume of diverse tasks. Each worker VMs runs a client process to serve its corresponding model. The requests are served as independent parallel threads to ensure timely predictions. We use \texttt{Python Sanic} web-server for communication with the master and worker VMs. Each worker VM runs tensorflow-serving~\cite{serving} to serve the inference requests. \\
\textbf{Load Balancer}: The master VMs runs a separate thread to monitor the importance sampling of all individual model pools. It keeps track of the number of requests served per model in the past 5 minutes. This information is used for calculating the weights per model for autoscaling decisions.  We integrate a \texttt{mongodb}~\cite{mongodb} database in the master node to maintain all information about  procured instances, spot-instance price list, and instance utilization. The load prediction model resides in the master VM which constantly records the arrival rate in adjacent windows. Recall that the details of the prediction were described in Section~\ref{sec:prediction}. The  DeepAREst~\cite{deepar} model was trained using \texttt{Keras}~\cite{keras} and \texttt{Tensorflow}, over 100 epochs with 2 layers, 32 neurons and a batch-size of 1. \\
\textbf{Model Cache}: We keep track of the model selected for ensembling on a per request constraint basis. The constraints are defined as \texttt{<latency,accuracy>} pair. The queries arriving with similar constraints can read the model cache to avoid re-computation for selecting the models. The model cache is implemented as a hash-map using \texttt{Redis}~\cite{redis} in-memory key-value store for fast access. \\
\textbf{Constraint specification}: {\color {black} We expose a simple API to developers, where they can specify the type of inference task (e.g., classification) along with the \texttt{<latency,accuracy>} constraints. Developers also need to indicate the primary objective between these two constraints. }
\emph{Cocktail} automatically chooses a set of single or ensemble models required to meet the developer specified constraints. \\
\noindent{\textbf{Discussion:}} Our accuracy and latency constraints are limited to the measurements from the available pretrained models. Note that changing the models or/and framework would lead to minor deviations. While providing latency and top-1\% accuracy of the pretrained models is an offline step in \emph{Cocktail}, we can calculate these values through one-time profiling and use them in the framework. All decisions related to VM autoscaling, bin-packing and load-prediction are reliant on the centralized mongodb database, which can become a potential bottleneck in terms of scalability and consistency. This can be mitigated by using fast distributed solutions like Redis~\cite{redis} and Zookeeper~\cite{zookeeper}. The DeepARest model is pre-trained using 60\% of the arrival trace. For varying load patterns, the model parameters can be updated by re-training in the background with new arrival rates.

\begin{table}
\centering
\footnotesize
\begin{tabular}{||c|c|c|c|c||}
\hline
\textbf{Dataset} & \textbf{Application} & \textbf{Classes} & \textbf{Train-set} & \textbf{Test-set} \\ \hline
ImageNet~\cite{imagenet} & Image & 1000 & 1.2M & 50K \\ \hline
CIFAR-100~\cite{cifar} & Image & 100 & 50K & 10K \\ \hline
SST-2~\cite{sst} & Text & 2 & 9.6K & 1.8K \\ \hline
SemEval~\cite{semeval} & Text & 3 & 50.3K & 12.2K \\ \hline
\end{tabular}%
\caption{{\color{black}Benchmark Applications and datasets.}}
\label{tbl:benchmarks}
\end{table}
\subsection{Evaluation Methodology} 
\label{sec:constraints}
We evaluate our prototype implementation on AWS EC2~\cite{c5} platforms. Specifically, we use \texttt{C5.xlarge, 2xlarge, 4xlarge, 8xlarge} for CPU instances and \texttt{p2.xlarge} for GPU instances. \\
\textbf{{Load Generator:}} We use different traces which are given as input to the load generator. Firstly, we use real-world request arrival traces from Wikipedia~\cite{wiki}, which exhibit typical characteristics of ML inference workloads as it has recurring diurnal patterns. The second trace is production twitter~\cite{twitter} trace which is bursty with unexpected load spikes. We use the first 1 hour sample of both the traces and they are scaled to have an average request rate of 50 req/sec.  \\
\textbf{{Workload:}} As shown in Table~\ref{tbl:benchmarks} we use  image-classification and Sentiment Analysis (text) applications with two datasets each for our evaluation. Sentiment analysis outputs the sentiment of a given sentence as positive negative and (or) neutral. We use 9 different prominently used text-classification models from transformers library~\cite{huggingface} (details available in appendix) designed using Google BERT~\cite{bert} architecture trained on \texttt{SST}~\cite{sst} and \texttt{SemEval}~\cite{semeval} dataset. Each request from the load-generator is modelled after a query with specific \texttt{<latency,accuracy>} constraints. The queries consist of images/text, which are randomly 
\begin{wrapfigure}{l}{0.25\textwidth}
\centering
\includegraphics[width=0.25\textwidth]{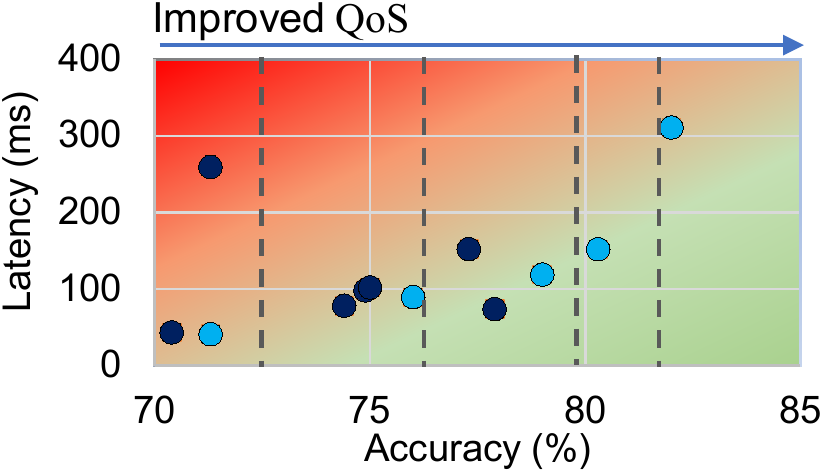}
\caption{Constraints used.}
\label{fig:constraints}
\end{wrapfigure}
picked from the test dataset. In our experiments, we use five different types of these constraints. {\color{black} As an example for  the \texttt{Imagenet} dataset shown in Figure~\ref{fig:constraints}, each constraint is a representative of <latency, accuracy> combination offered by single models (shown in Table~\ref{tbl:models}). We use one constraint (blue dots) each from five different regions (categorized by dotted lines) picked in the increasing order of accuracy. The ensemble size for these constraints ranges from small (2) to large (10), as shown in Table~\ref{tbl:ensemble}. Note that the latency is the raw model execution latency, and does not include the additional network-transfer overheads incurred. We picked the constraints using a similar procedure for \texttt{CIFAR-100}, \texttt{SST-2} and \texttt{SemEval} (twitter tweets) datasets as well.} 
We model two different workload mixes by using a combination of these five query constraint types. Based on the decreasing order of {accuracy}, we categorize them into \emph{Strict} and \emph{Relaxed} workloads. 
\begin{figure*}
\begin{minipage}{0.99\linewidth}
\begin{subfigure}[t]{.24\textwidth}
\centering
\includegraphics[width=0.9\textwidth]{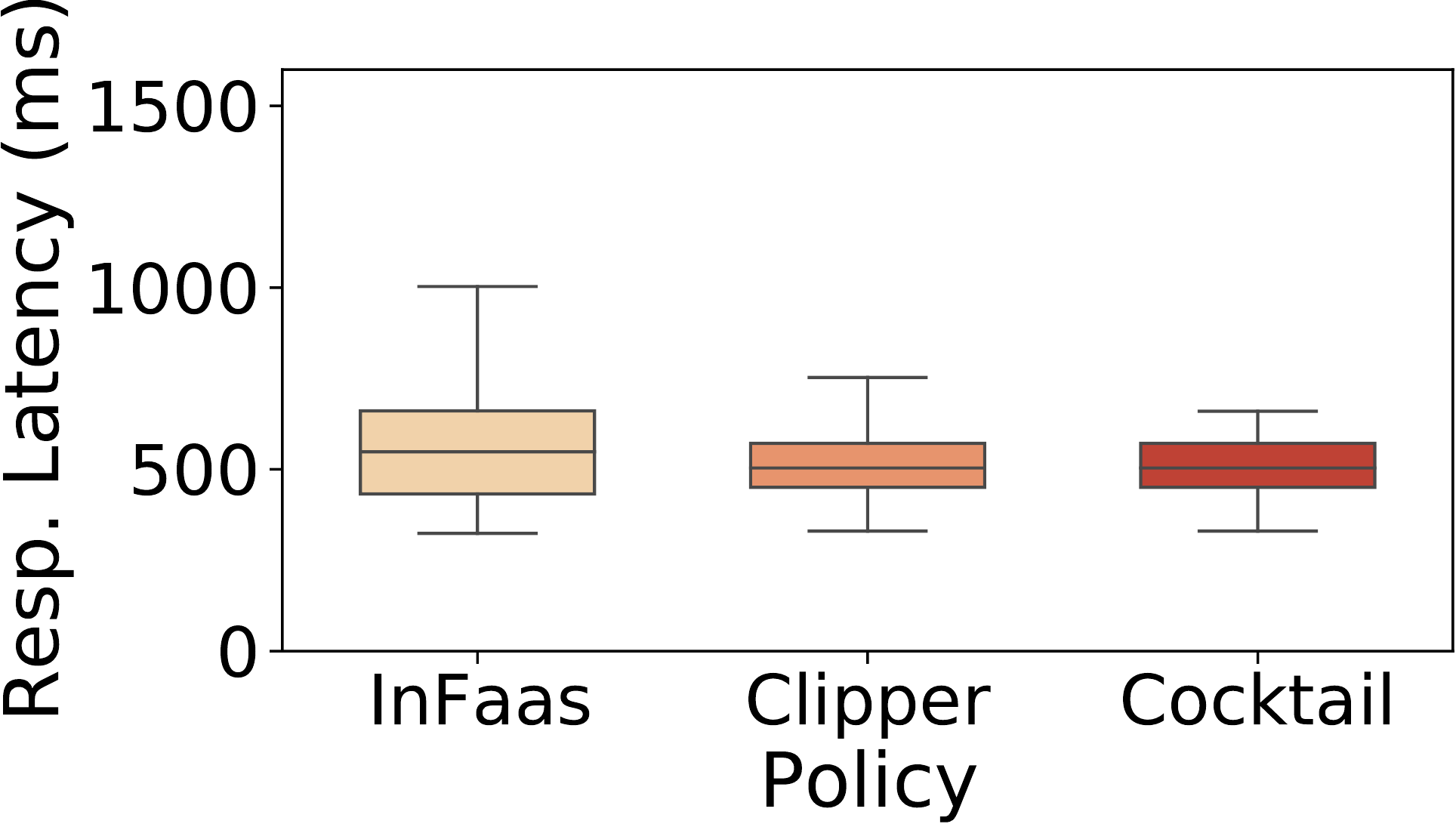}
\caption{Wiki-trace: \emph{Strict} workload.}
\label{fig:tail1-wiki}
\end{subfigure}
\begin{subfigure}[t]{0.24\textwidth}
\centering
\includegraphics[width=0.9\textwidth]{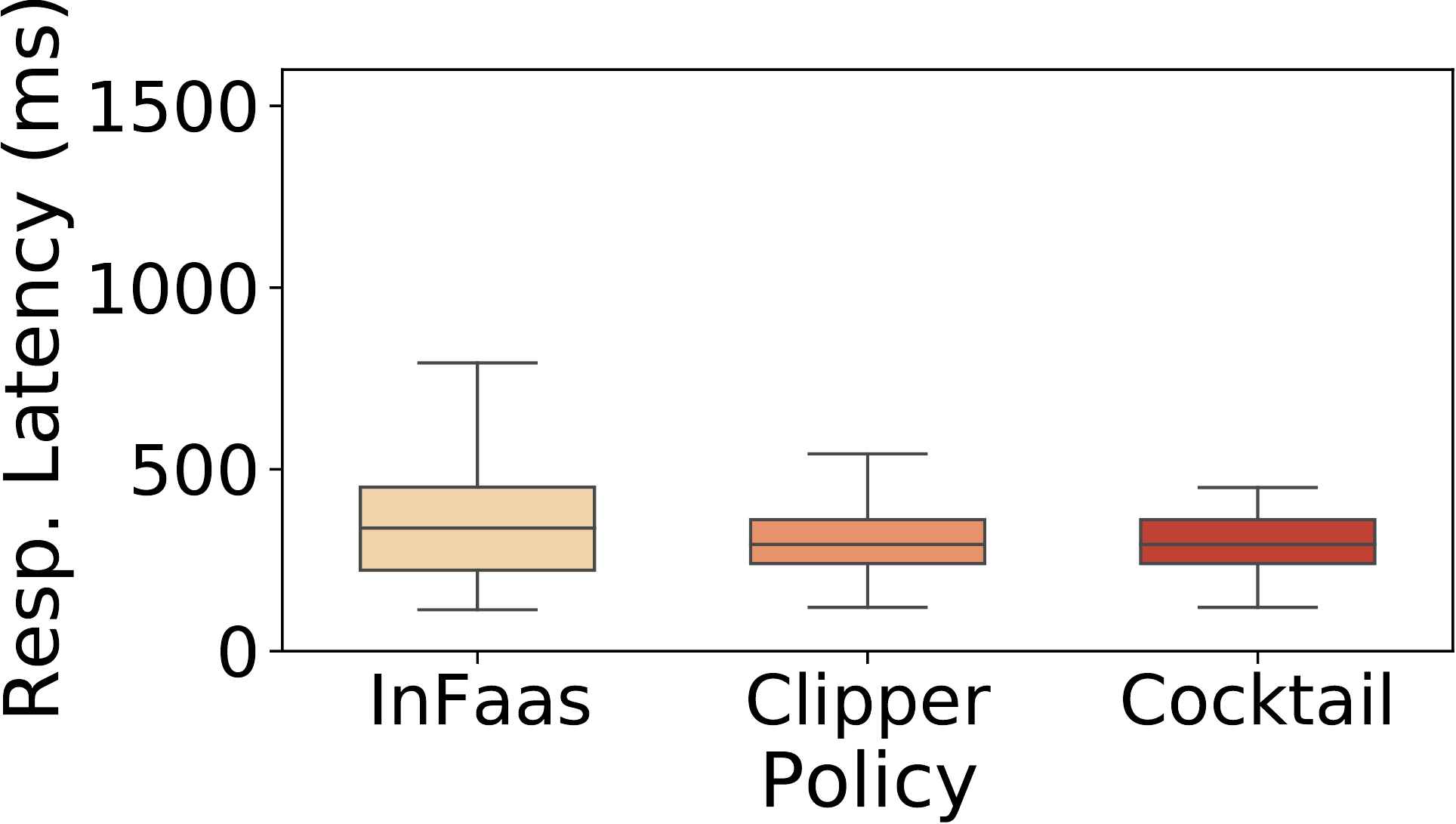}
\caption{Wiki-trace: \emph{Relaxed} workload.}
\label{fig:tail2-wiki}
\end{subfigure}
\begin{subfigure}[t]{.24\textwidth}
\centering
\includegraphics[width=0.9\textwidth]{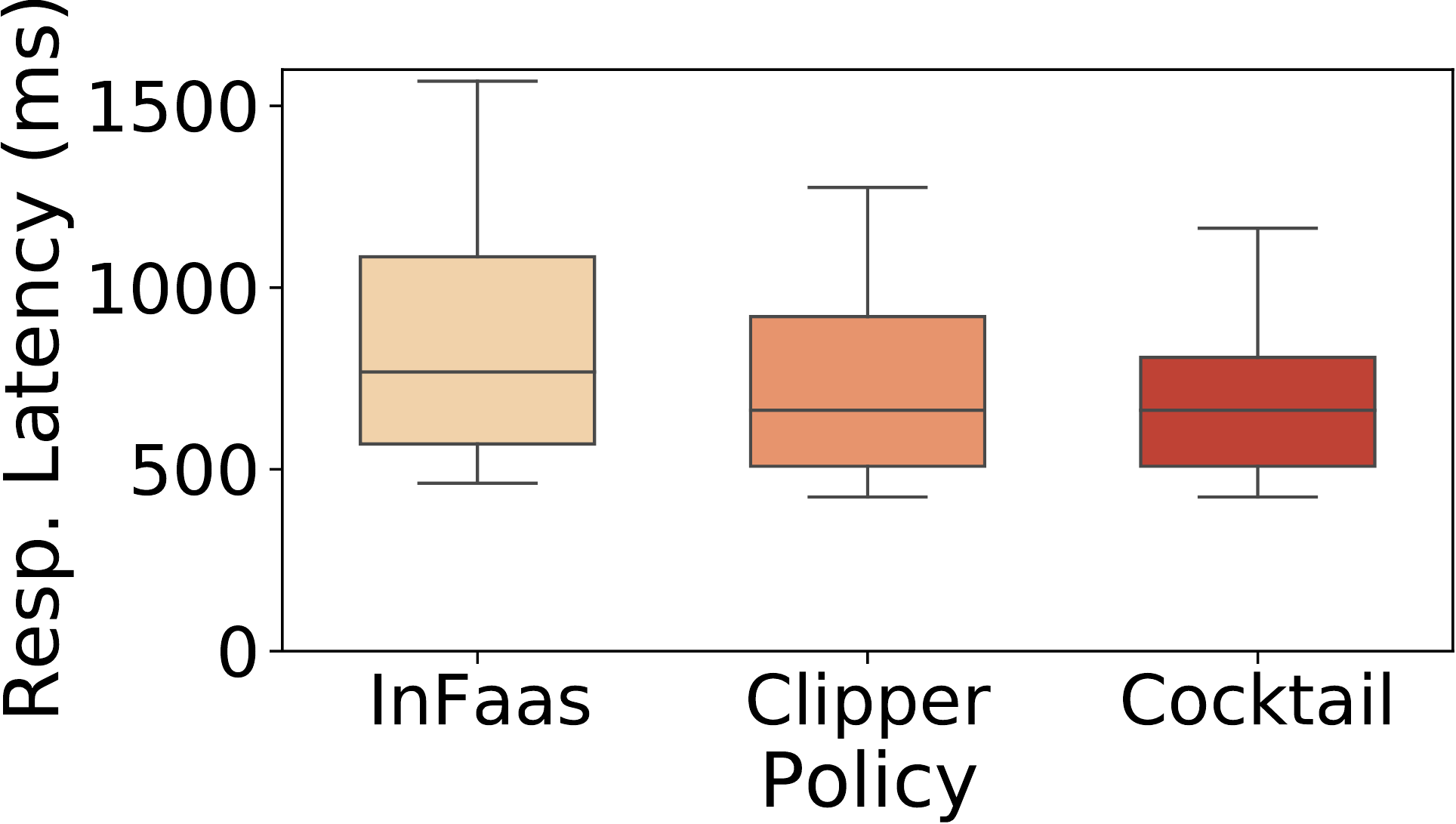}
\caption{Twitter-trace: \emph{Strict} workload.}
\label{fig:tail1-twitter}
\end{subfigure}
\begin{subfigure}[t]{0.24\textwidth}
\centering
 \includegraphics[width=0.9\textwidth]{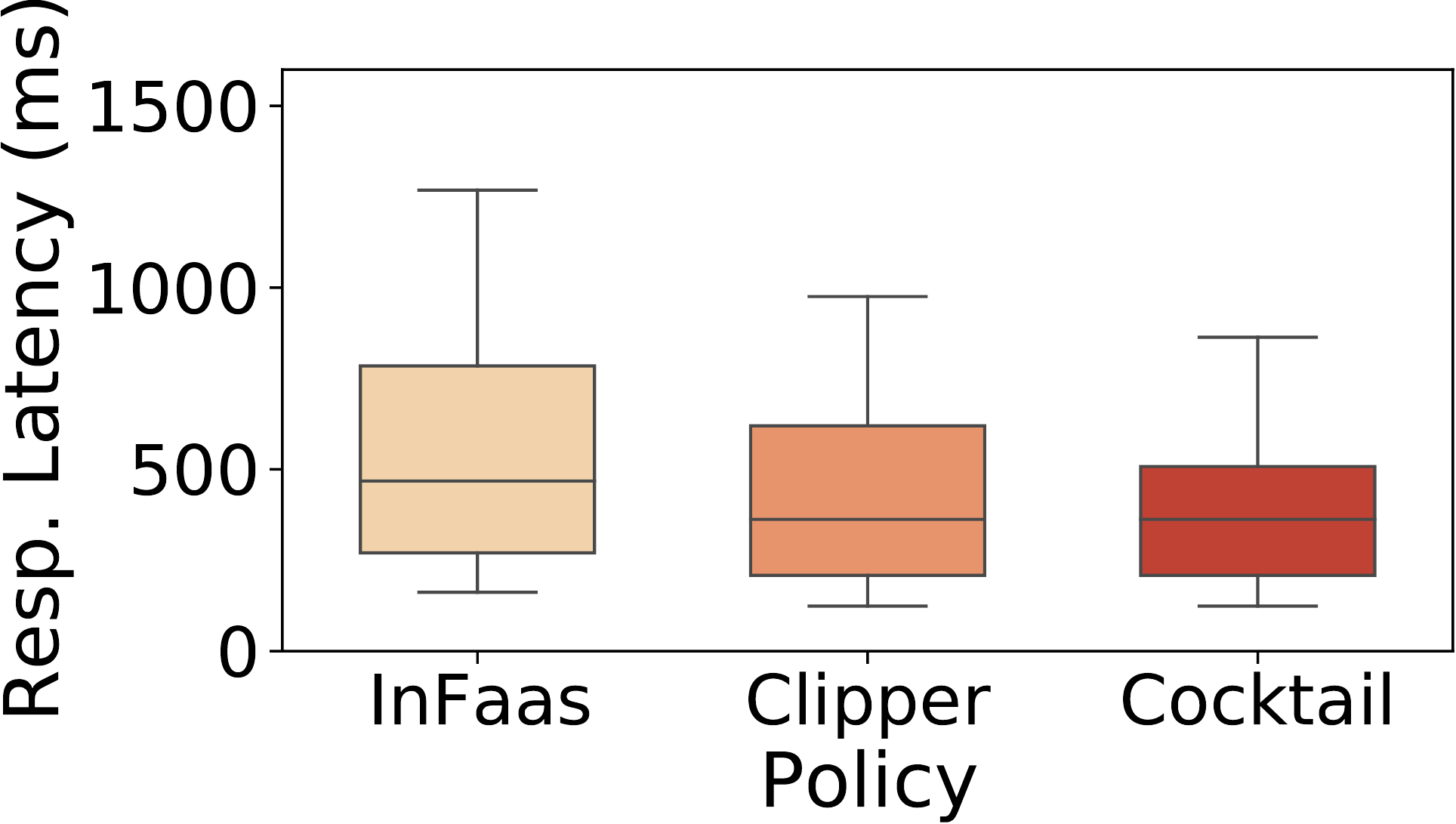}
\caption{Twitter-trace: \emph{Relaxed} workload.}
\label{fig:tail2-twitter}
\end{subfigure}
\end{minipage}
\caption{Latency Distribution of \emph{InFaas}, \emph{Clipper} and \emph{Cocktail} for two workload mixes using both Wiki and Twitter traces.}
\label{fig:tail}
\end{figure*}
\subsubsection{Evaluation Metrics}
\label{sec:eval}
{\color{black}Most of our evaluations of \emph{Cocktail} for image-classification are performed using the \texttt{Imagenet} dataset. To further demonstrate the sensitivity of Cocktail to dataset and applicability to other classification applications, we also evaluate it using \texttt{CIFAR-100} and Sentiment-Analysis application.} We use three important metrics: response latency, cost and accuracy for evaluating and comparing our design to other state-of-the-art systems. The response latency metric includes model inference latency, communication/network latency and synchronization overheads. Queries that do not meet response latency requirements (>700ms) are considered as SLO violations. The cost metric is the billing cost from AWS, and the accuracy metric is measured as the percentage of requests that meet the target accuracy requirements. We compare these metrics for \emph{Cocktail} against (i) \emph{InFaas}~\cite{infaas}, which is our baseline that employs single model selection policy; (ii) {\color {black}\emph{Clipper}~\cite{clipper},  which uses static full model selection policy (analogous to AWS AutoGluon); and (iii) \emph{Clipper-X} which is an enhancement to \emph{Clipper} with a simple model selection (drop one model at a time) that does not utilize the \textit{mode}-based policy enforced in \emph{Cocktail}}. Both \emph{InFaas} and \emph{Clipper} share \emph{Cocktail}'s implementation setup to ensure a fair comparison with respect to our design and execution environment. For instance, both \emph{Clipper} and \emph{InFaas} employ variants of a reactive autoscaler as described in Section~\ref{sec:autoscaler}. However, in our setup, both benefit from the distributed autoscaling and prediction policies, thus eliminating variability. Also note that \emph{InFaas} is deployed using OnDemand instances, while both \emph{Clipper} and \emph{Cocktail} use spot instances. In our evaluations, we use {\em accuracy} as the primary parameter. As a result, all ensemble requests will wait until they receive a response from their respective model workers.


\section{Analysis of Results}
\label{sec:results}
This section discusses the experimental results of \emph{Cocktail} using the Wiki and Twitter traces.
\subsection{Latency, Accuracy and Cost Reduction}
\textbf{Latency Distribution}:
Figure~\ref{fig:tail} shows the distribution of total response latency in a standard box-and-whisker plot. 
The boundaries of the box-plots depict the 1st quartile (25th percentile (PCTL)) and 3rd quartile (75th PCTL), the whiskers plot the minimum and maximum (tail) latency and the middle line inside the box depict the median (50 PCTL). The total response latency includes additional 200-300ms incurred for query serialization and data transfer over network. It can be seen that the maximum latency of \emph{Cocktail} is similar to the 75th PCTL latency of \emph{InFaas}. This is because the single model inference have up to 2x higher latency to achieve higher accuracy. Consequently, this leads to 35\% SLO violations for \emph{InFaas} in the case of Strict workload. In contrast, both \emph{Cocktail} and \emph{Clipper} can reach the accuracy at lower latency due to ensembling, thus minimizing SLO violations to 1\%. Also, the tail latency is higher for Twitter trace (Figure~\ref{fig:tail1-twitter}, \ref{fig:tail2-twitter}) owing to its bursty nature. Note that the tail latency of \emph{Clipper} is still higher than \emph{Cocktail} because \emph{Clipper} ensembles more models than \emph{Cocktail}, thereby resulting in straggler tasks in the VMs. The difference in latency between \emph{Cocktail} and \emph{InFaas} is lower for \emph{Relaxed} workload when compared to \emph{Strict} workload (20\% lower in tail). Since the \emph{Relaxed} workload has much lower accuracy constraints, smaller models are able to singularly achieve the accuracy requirements at lower latency.\\
\textbf{Accuracy violations}: Table~\ref{tbl:accuracy} plots the percentage of queries that meet the target  accuracy. The accuracy is measured as a moving window average with size 200 for all the requests in the workload. \begin{wraptable}{r}{0.25\textwidth}
\footnotesize
\centering
\begin{tabular}{|l|l|l|}
\hline
& \multicolumn{2}{l|}{{Accuracy Met (\%)}} \\ \cline{2-3} 
\multirow{-2}{*}{\textbf{Scheme}} & Strict & Relaxed \\ \hline
\emph{InFaas} & 21 & 71 \\ \hline
\emph{Clipper} & 47 & 89 \\ \hline
\emph{Cocktail} & 56 & 96 \\ \hline
\end{tabular}
\caption{Requests meeting target accuracy averaged for both Trace.}
\label{tbl:accuracy}
\end{wraptable}It is evident both \emph{Clipper} and \emph{Cocktail} can meet the accuracy for 56\% of requests, which is 26\% and 9\% more than \emph{InFaas} and \emph{Clipper} respectively. This is because, intuitively ensembling leads to higher accuracy than single models. However, \emph{Cocktail} is still 9\% better than \emph{Clipper} because the class-based weighted voting, is efficient in breaking ties when compared to weighting averaging used in \emph{Clipper}. {\color{black} Since majority voting can include  
ties in votes, we analyzed the number of ties, which were correctly predicted for all the queries. \emph{Cocktail} was able to deliver correct predictions for 35\% of the tied votes, whereas breaking the ties according to \emph{Clipper}'s policy led only to 20\% correct predictions.} 
\begin{figure}
\begin{minipage}{0.99\linewidth}
\begin{subfigure}{.48\textwidth}
\centering
\includegraphics[width=0.99\textwidth]{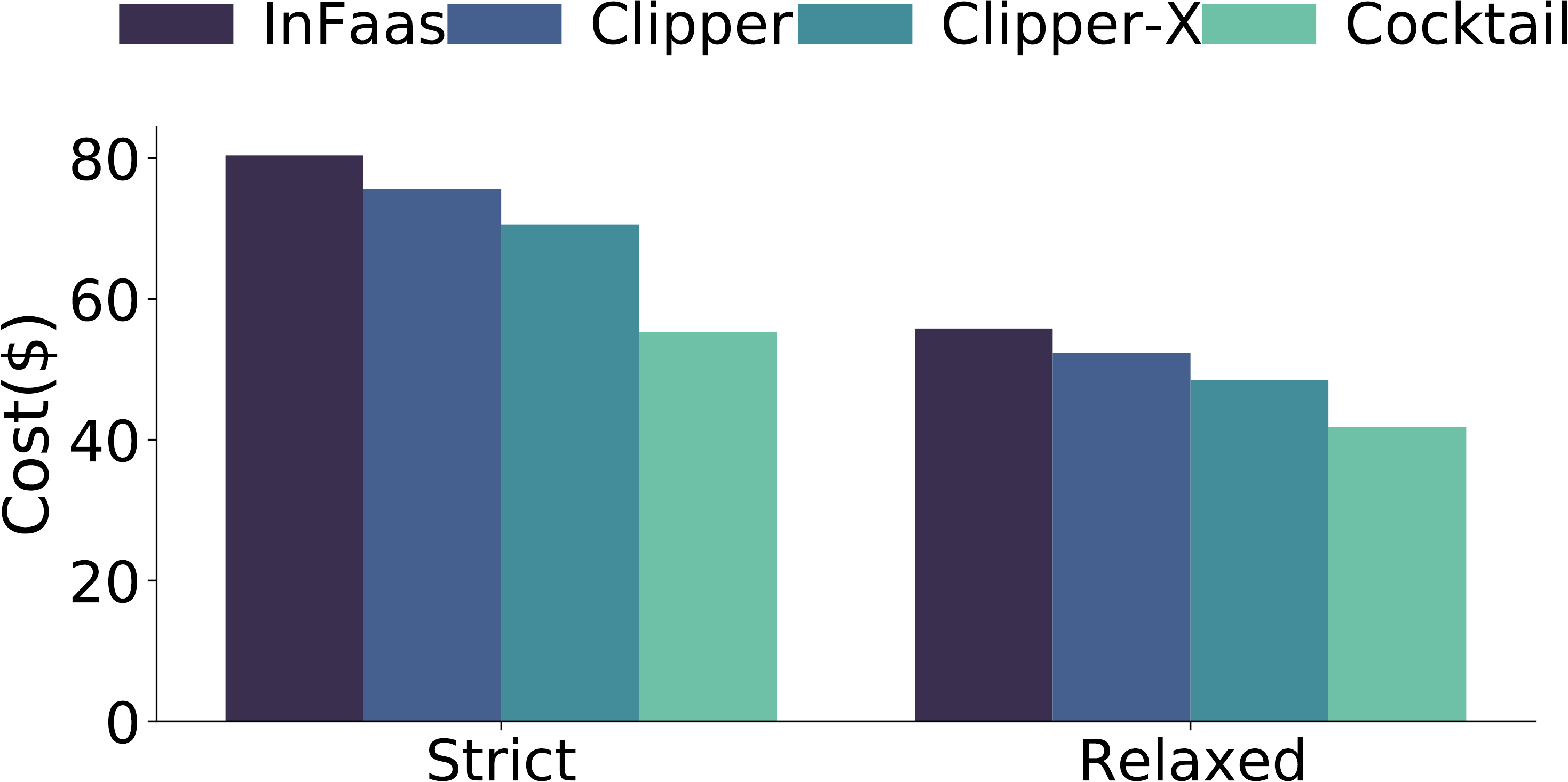}
\caption{Wiki Trace.}
\label{fig:wiki-cost}
\end{subfigure}
\begin{subfigure}{0.48\textwidth}
\centering
 \includegraphics[width=0.99\textwidth]{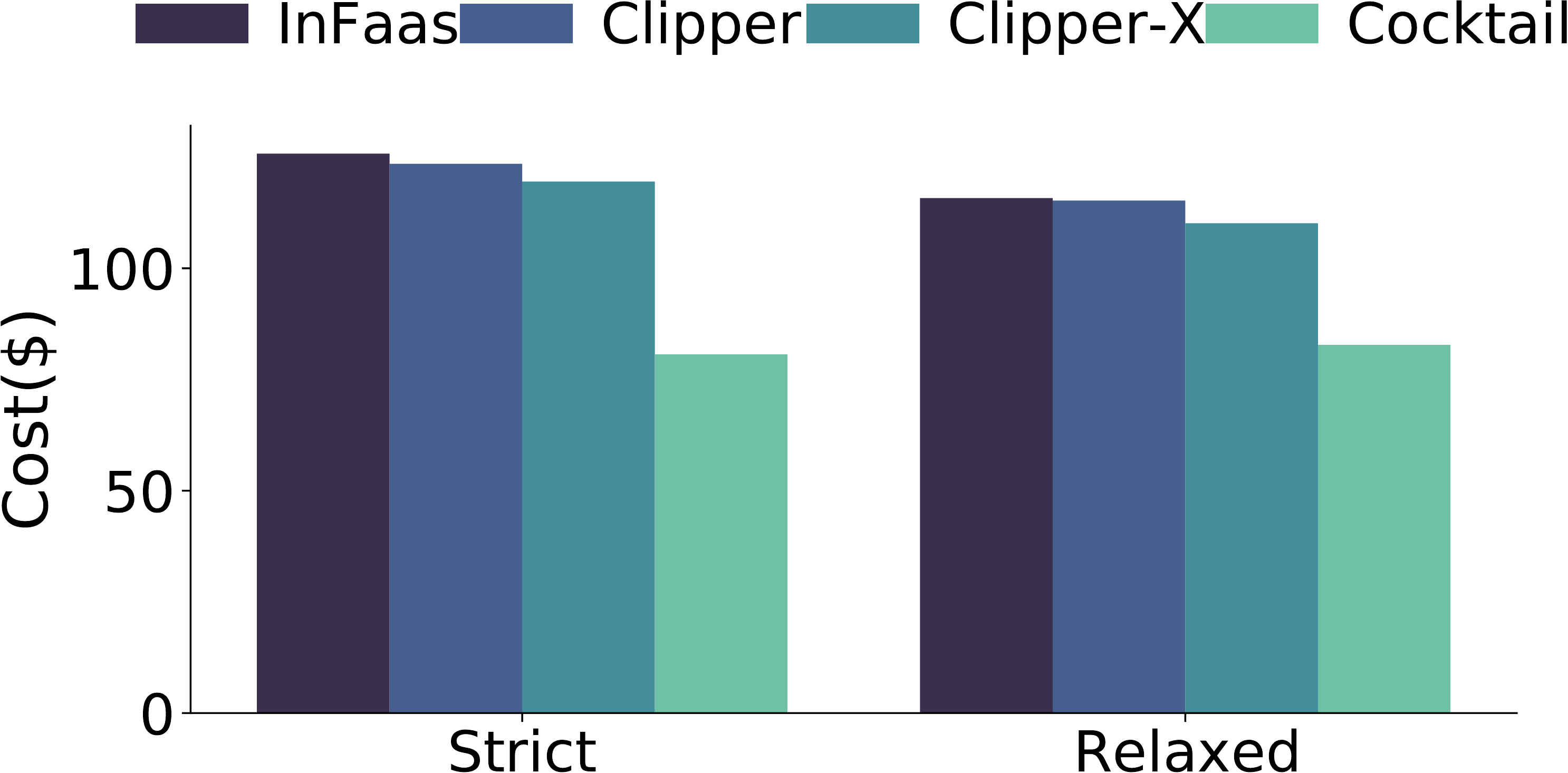}
\caption{Twitter Trace.}
\label{fig;tweet-cost}
\end{subfigure}
\end{minipage}
\caption{Cost savings of \emph{Cocktail} compared to three schemes.}
\label{fig:cost}
\end{figure}
Note that, changing the target accuracy to tolerate a 0.5\% loss,  increases the percentage of requests that meet accuracy to 81\% for \emph{Cocktail}, when compared to 61\% for \emph{InFaas}. The requests meeting accuracy are generally higher for the \emph{Relaxed} workload because the target accuracy is much lower. {\color{black} Overall, \emph{Cocktail} was able to deliver an accuracy of 83\% and 79.5\% on average for the \emph{Strict} and \emph{Relaxed} workloads, respectively. This translates to 1.5\% and 1\% better accuracy than \emph{Clipper} and \emph{InFaas}.}  We do not plot the results for \emph{Clipper-X}, which achieves similar accuracy to \emph{Cocktail}, but uses more models as explained in Section~\ref{sec:model-selection-benefits}.\\
\textbf{Cost Comparison:}
Figure~\ref{fig:cost} plots the cost savings of \emph{Cocktail} when compared to \emph{InFaas}, \emph{Clipper} and \emph{Clipper-X} policies. It can be seen that, \emph{Cocktail} is up to 1.45$\times$ more cost effective than \emph{InFaas} for \emph{Strict} workload. In addition, \emph{Cocktail} reduces cost by  1.35$\times$ and 1.27$\times$ compared to \emph{Clipper} and \emph{Clipper-X} policies, owing to its dynamic model selection policy, which minimizes the resource footprint of ensembling. On the other hand, \emph{Clipper} uses all models in ensemble and the \emph{Clipper-X} policy does not right size the models as aggressively as \emph{Clipper}, hence they are more expensive. 
Note that, all the schemes incur higher cost for twitter trace (Figure~\ref{fig;tweet-cost}) compared to wiki trace (Figure~\ref{fig:wiki-cost}). This is because the twitter workload is bursty, thereby leading to intermittent over-provisioned VMs. 

\begin{figure}
\begin{minipage}{0.99\linewidth}
\centering
\begin{subfigure}[t]{0.44\textwidth}
\centering
 \includegraphics[width=0.9\textwidth]{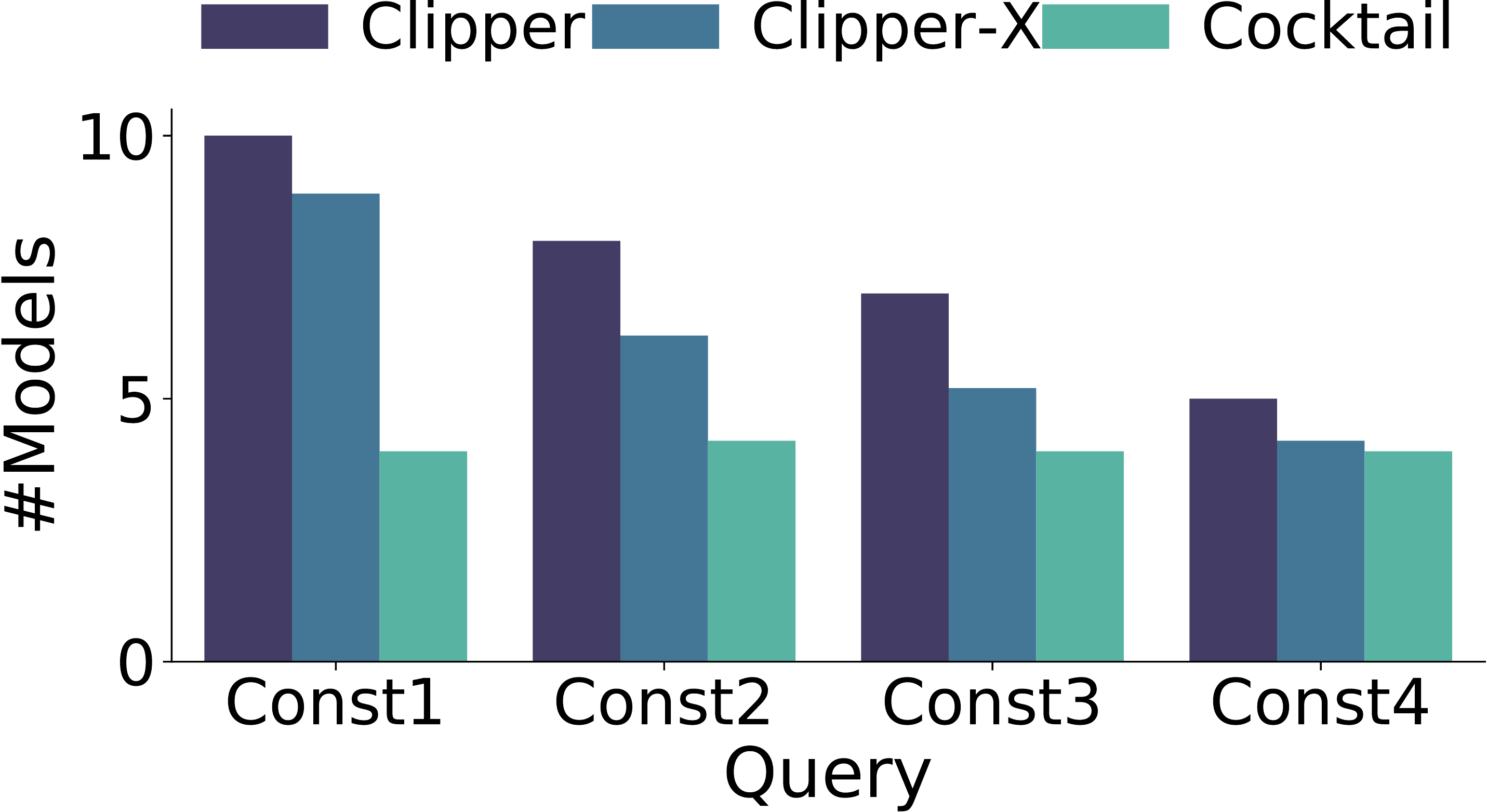}
\caption{Average number of models used in the ensemble.}
\label{fig:model-reduced}
\end{subfigure}
\begin{subfigure}[t]{0.44\textwidth}
\centering
 \includegraphics[width=0.9\textwidth]{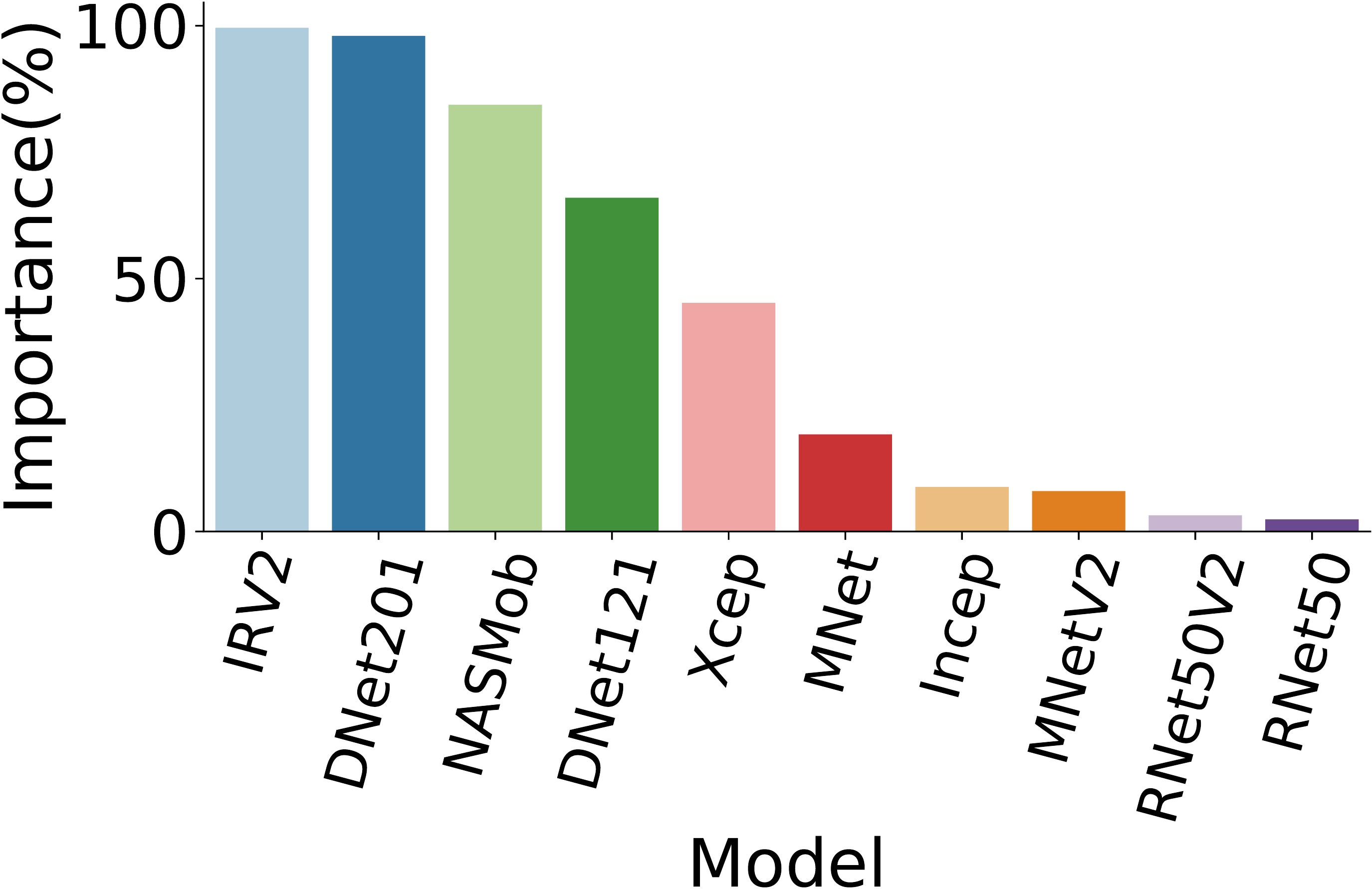}
\caption{Distribution of requests served by each individual model.}
\label{fig:model-breakdown}
\end{subfigure}
\end{minipage}
\caption{Benefits of dynamic model selection policy.}
\end{figure}
\begin{figure*}
\begin{minipage}[t]{0.99\linewidth}
\centering
\begin{subfigure}[t]{0.24\textwidth}
\centering
 \includegraphics[width=0.99\textwidth]{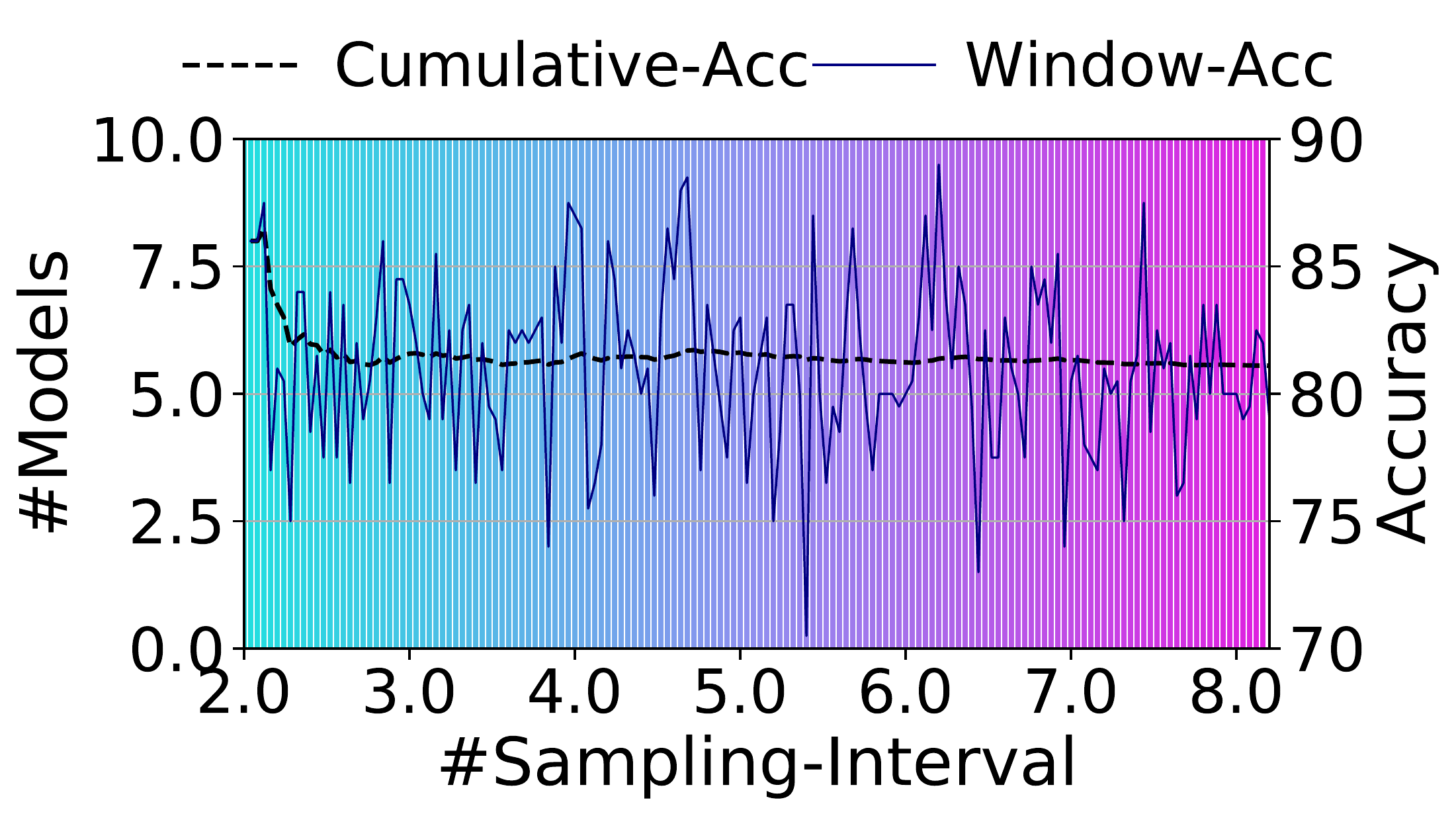}
\caption{\emph{Clipper}}
\label{fig:model3}
\end{subfigure}
\begin{subfigure}[t]{0.24\textwidth}
\centering
 \includegraphics[width=0.99\textwidth]{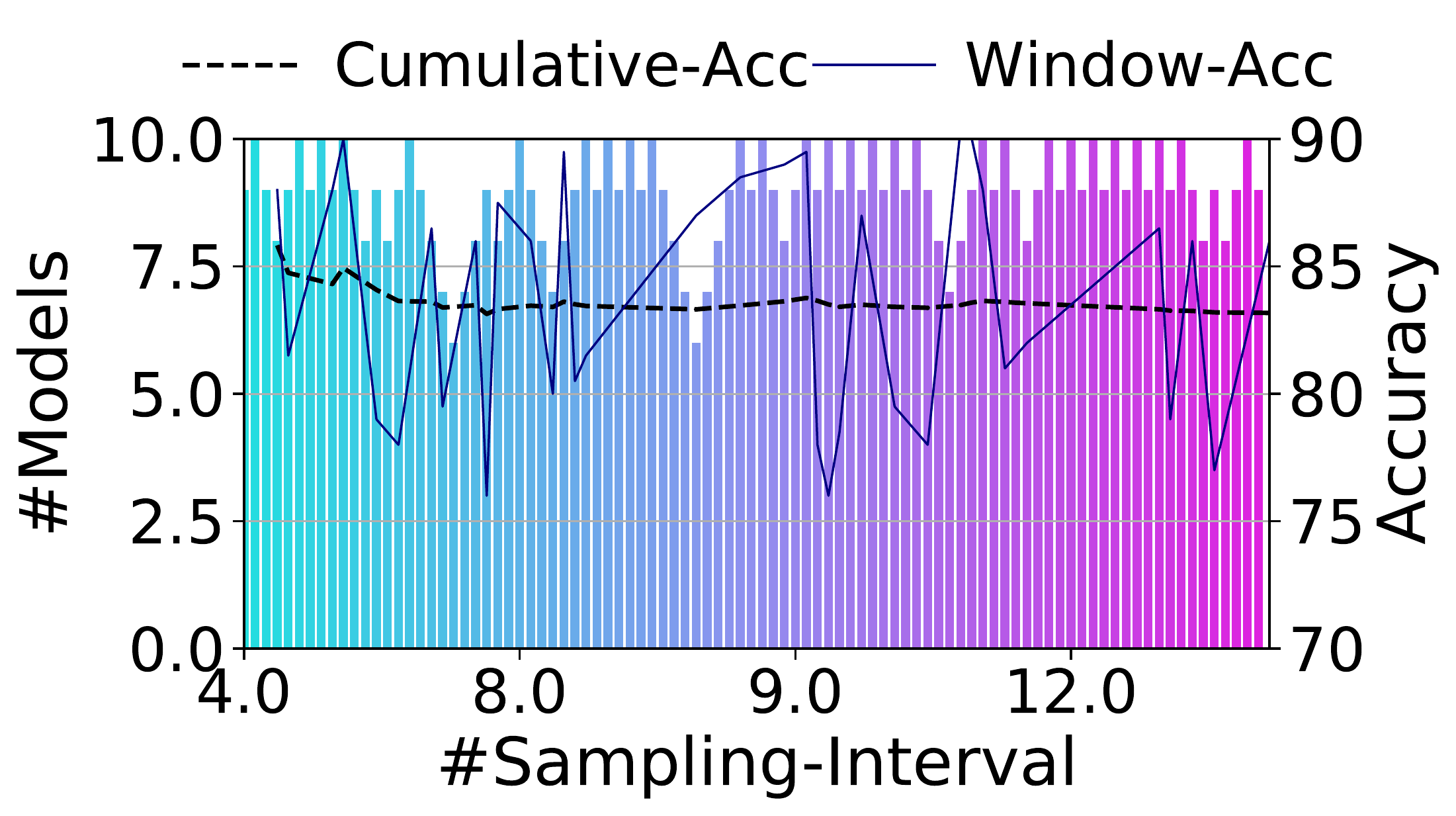}
\caption{\emph{Clipper-X}}
\label{fig:model2}
\end{subfigure}
\begin{subfigure}[t]{.24\textwidth}
\centering
\includegraphics[width=0.99\textwidth]{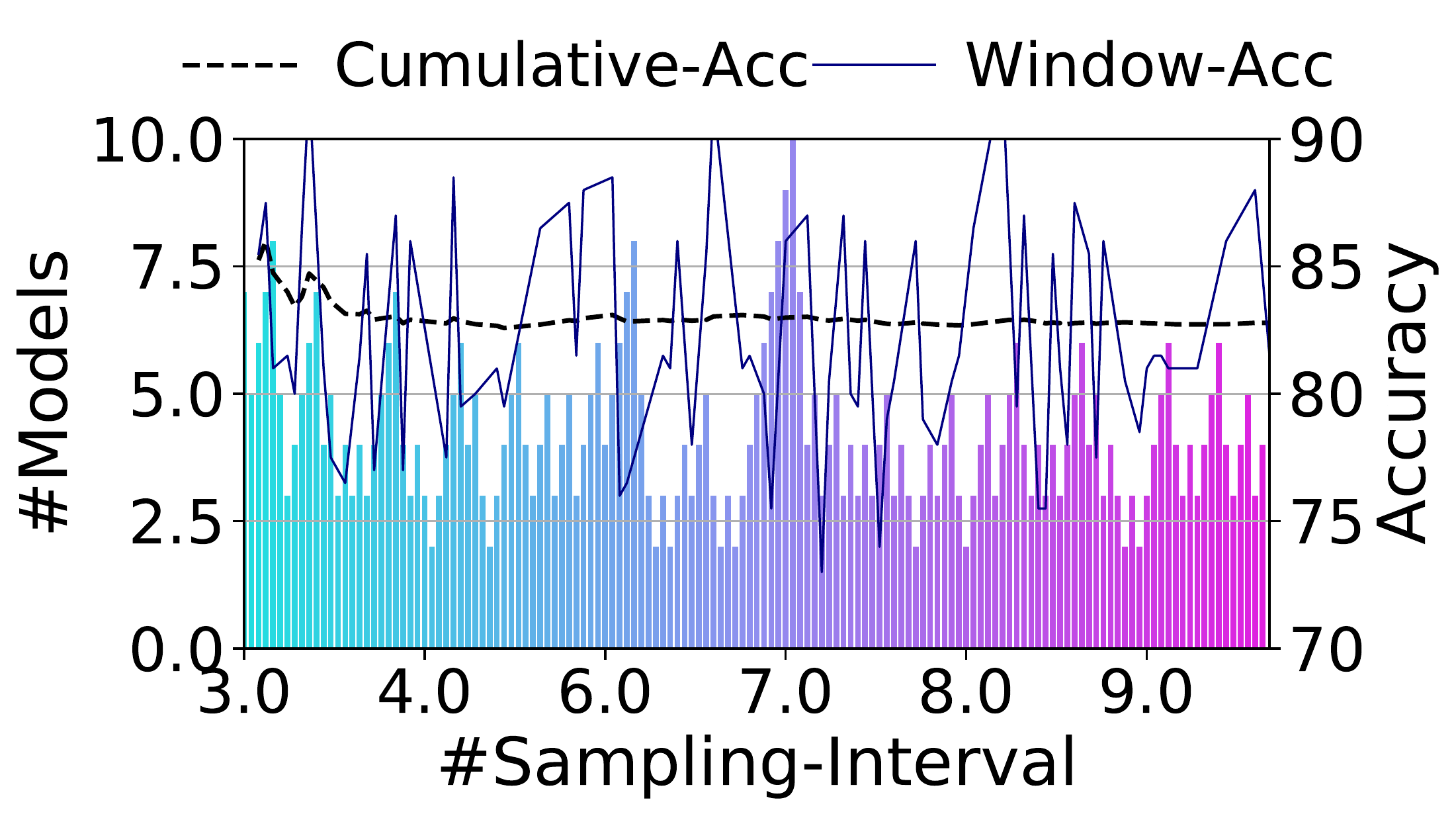}
\caption{\emph{Cocktail}}
\label{fig:model1}
\end{subfigure}
\begin{subfigure}[t]{0.24\textwidth}
\centering
 \includegraphics[width=0.99\textwidth]{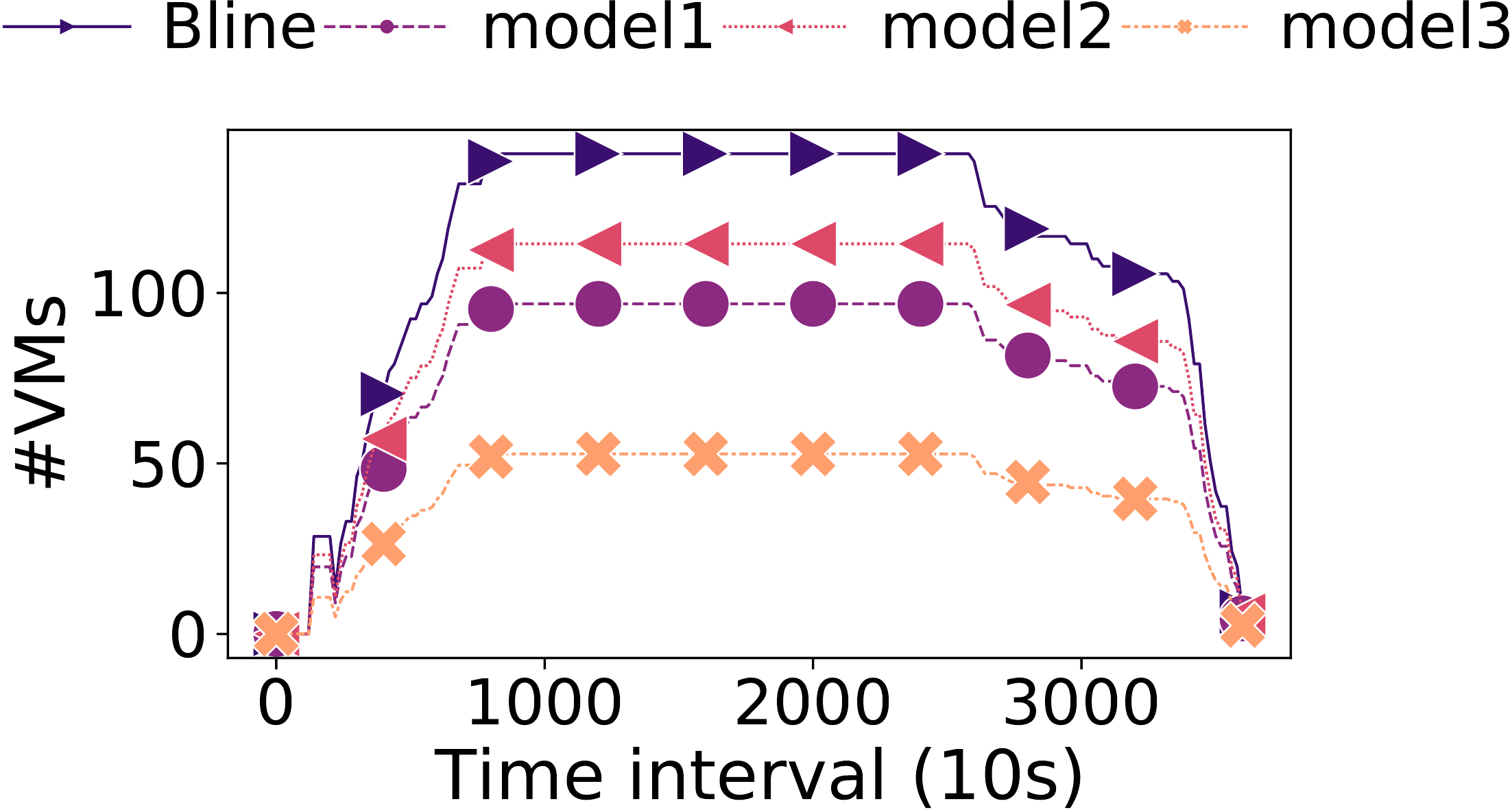}
\caption{Cumulative \#VMs over time.}
\label{fig:VMs}
\end{subfigure}
\end{minipage}
\caption{Figures (a), (b) and (c) shows the number of models used in ensemble with corresponding cumulative accuracy and window accuracy over a 1 hour period for requests under const-1. Figure (d) shows the effects of distributed autoscaling with importance sampling.}
\label{fig:models-select}
\end{figure*}
\subsection{Key Sources of Improvements}
{The major improvements in terms of cost, latency, and accuracy} in \emph{Cocktail} are explained below. For brevity in explanation, the results are averaged across Wiki and Twitter traces for strict workload. 
\subsubsection{Benefits from dynamic model selection}
\label{sec:model-selection-benefits}
{Figure~\ref{fig:model-reduced} plots the average number of} models used for queries falling under the first four different constraint (const) types. Here, \emph{Cocktail} reduces the number of models by up to 55\% for all four query types. This is because our dynamic policy ensures that the number of models are well within $N/2$ most of the time, whereas the \emph{Clipper-X} policy does not aggressively scale down models. \emph{Clipper}, on the other hand, is static and always uses all the models. The percentage of model-reduction is lower for const-2, 3 and 4 because, the total models used in the ensemble is less than const-1 (8, 7 and 6 models, respectively). Still, the savings in terms of cost will be significant because even removing one model from the ensemble amounts to $\sim$20\% cost savings in the long run (\emph{Clipper} vs \emph{Clipper-X} ensemble in Figure~\ref{fig:cost}). Thus, the benefits of \emph{Cocktail} are substantial for large ensembles while reducing the number of models for medium-sized ensembles as well.

Figure~\ref{fig:model-breakdown} shows the breakdown of the percentage of requests (const-1) served by the each model. As seen, InceptionResNetV2, Densenet-201, Densenet121, NasnetMobile and Xception are the top-5 most used models in the ensemble. Based on Table~\ref{tbl:models}, if we had statically taken the top $N/2$ most accurate models, NasNetmobile would not have been included in the ensemble. However, based on the input images sent in each query, our model selection policy has been able to identify NasNetMobile to be a significantly contributing model in the ensemble. Further, the other 5 models are used by up to 25\% of the images. Not including them in the ensemble would have led to severe loss in accuracy. {\color{black} But, our dynamic policy with the class-based weighted voting, adapts to input images in a give interval by accurately selecting the best performing model for each class}.  To further demonstrate the effectiveness of our dynamic model selection, Figure~\ref{fig:model3},~\ref{fig:model2},\ref{fig:model1} plots the number models in every sampling interval along with cumulative accuracy and window accuracy within each sampling interval for three schemes. We observe that \emph{Cocktail} can effectively scale up and scale down the models while maintaining the cumulative accuracy well within the threshold. More than 50\% of the time the number of models are maintained between 4 to 5, because the dynamic policy is quick in detecting accuracy failures and recovers immediately by scaling up models. However, \emph{Clipper-X} does not scale down models as frequently as \emph{Cocktail}, while ensuring similar accuracy. \emph{Clipper} is less accurate than \emph{Cocktail} and further it uses all 10 models throughout.  
\begin{figure}
\begin{minipage}[t]{0.99\linewidth}
\begin{subfigure}[t]{.45\textwidth}
\centering
\includegraphics[width=0.99\textwidth]{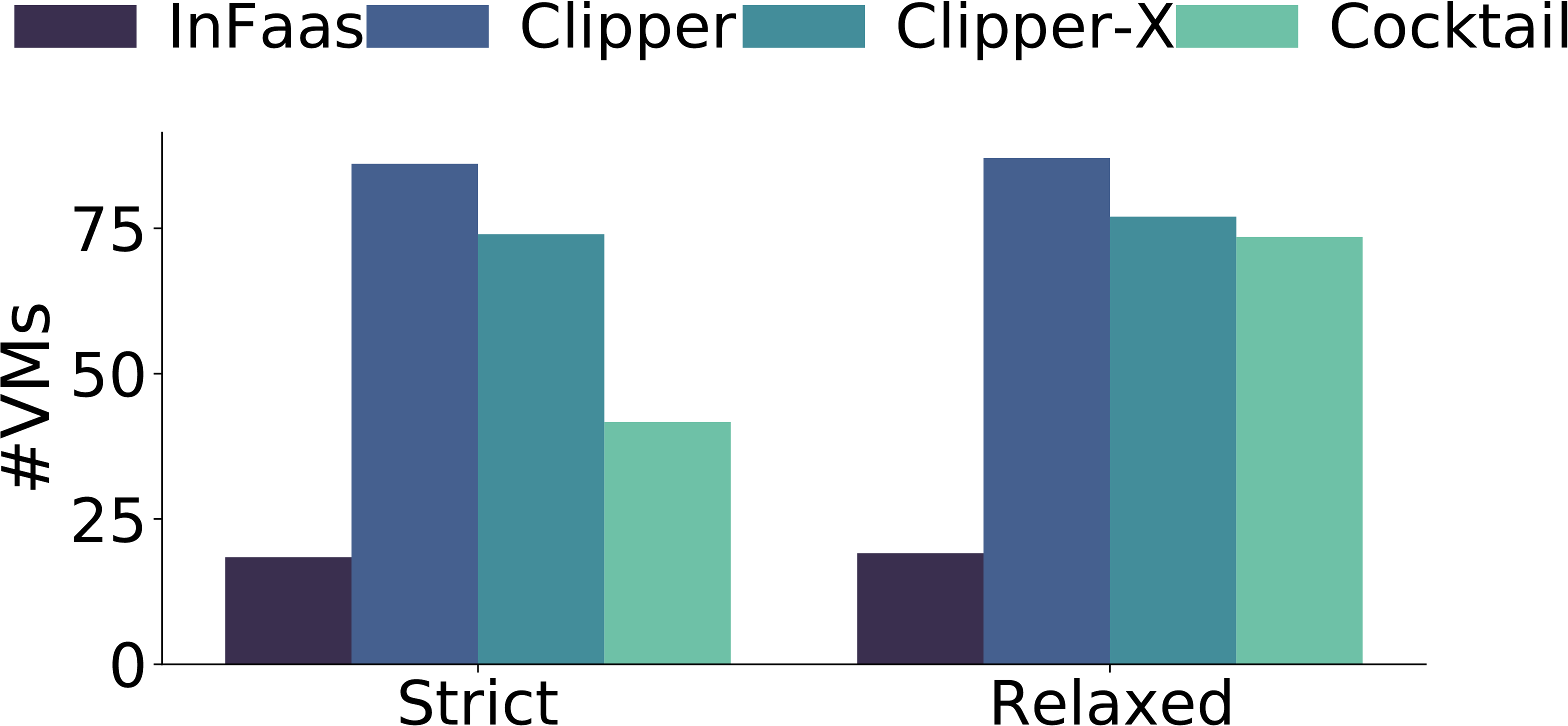}
\caption{Wiki Trace.}
\label{fig:wiki-vm}
\end{subfigure}
\begin{subfigure}[t]{0.45\textwidth}
\centering
\includegraphics[width=0.99\textwidth]{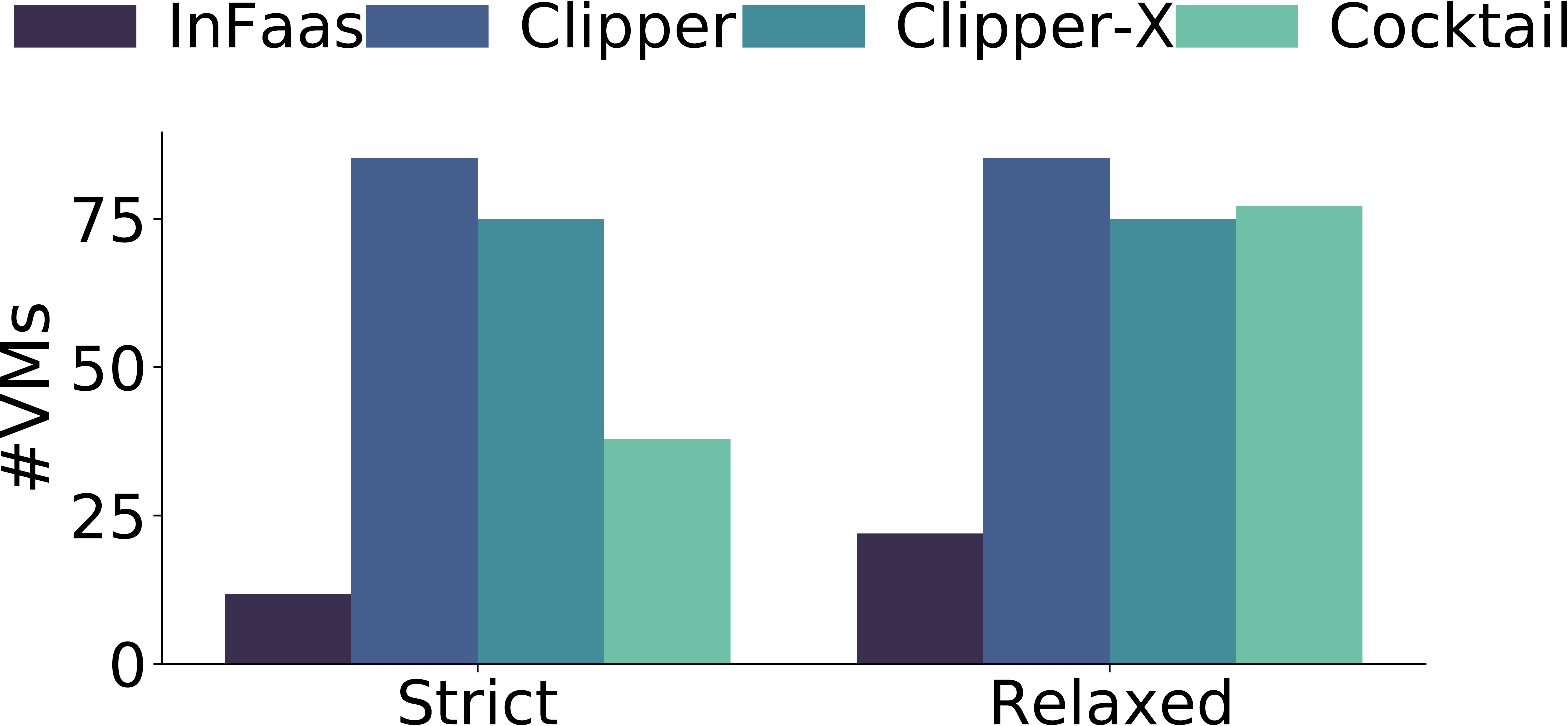}
\caption{Twitter Trace.}
\label{fig;tweet-vm}
\end{subfigure}
\end{minipage}
\caption{Number of VMs spawned for all four schemes.}
\label{fig:num-VMs}
\end{figure}
\subsubsection{Benefits from Autoscaling}
Figure~\ref{fig:num-VMs} plots the reduction in the number of VMs used by all four schemes.
It can be seen that both \emph{Cocktail} and \emph{Clipper-X} spawn 49\% and 20\% fewer VMs than \emph{Clipper} for workload-1 on Twitter trace. \emph{Cocktail} spawns 29\% lesser VMs on top of \emph{Clipper-X}, because it is not aggressive enough like \emph{Cocktail} to downscale more models at every interval. It is to be noted that the savings are lower for \emph{Relaxed} workload because, the number of models in the ensemble are inherently low, thus leading to reduced benefits from scaling down the models. Intuitively, \emph{InFaas} has the least number of VMs spawned because it does not ensemble models. \emph{Cocktail} spawns upto 50\% more VMs than \emph{InFaas}, but in turns reduces accuracy loss by up to 96\% (shown in Table~\ref{tbl:accuracy})

To further capture the benefits of the weighted autoscaling policy, Figure~\ref{fig:VMs} plots the number of VMs spawned over time for the top-3 most used models in the ensemble for const-1. The Bline denotes number of VMs that would be spawned without applying the weights. Not adopting an importance sampling based weighted policy would result in equivalent number of VMs as the Bline for all models. However, since \emph{Cocktail} exploits importance sampling by keeping track of the frequency in which models are selected, the number of VMs spawned for model1, model2 and model-3 is upto 3$\times$ times lesser than uniform scaling. Figure~\ref{fig:model-breakdown} shows the most used models in decreasing order of importance. The autoscaling policy effectively utilizes this importance factor in regular intervals of 5 minutes. Despite using multiple models for a single inference, importance sampling combined with aggressive model pruning, greatly reduces the resource footprint  which directly translates to the cost savings in \emph{Cocktail}. \vspace{-1mm}
\subsubsection{Benefits of Transient VMs}
The cost-reductions in \emph{Cocktail} are akin to cost-savings of transient VMs compared to On-Demand (OD) VMs.  
We profile the spot price of 4 types of \texttt{C5} EC2 VMs over a 2-week period in August 2020. It was seen that, the spot instance prices have predictable fluctuations. When compared to the OD price 
, they were up to 70\% cheaper. This price gap is capitalized in \emph{Cocktail} to reduce the cost of instances consumed by ensembling. Note that, we set the bidding price conservatively to 40\% of OD. Although, \emph{Cocktail} spawns about 50\% more VMs than \emph{InFaas}, the high $P_f$ of small models and spot-instance price reductions combined with autoscaling policies lead to the overall 30-40\% cost savings. 

\subsection{Sensitivity Analysis}
\begin{wrapfigure}[19]{l}{0.23\textwidth}
\centering
\begin{minipage}{0.9\linewidth}
\centering
\begin{subfigure}{0.99\textwidth}\includegraphics[width=0.99\textwidth]{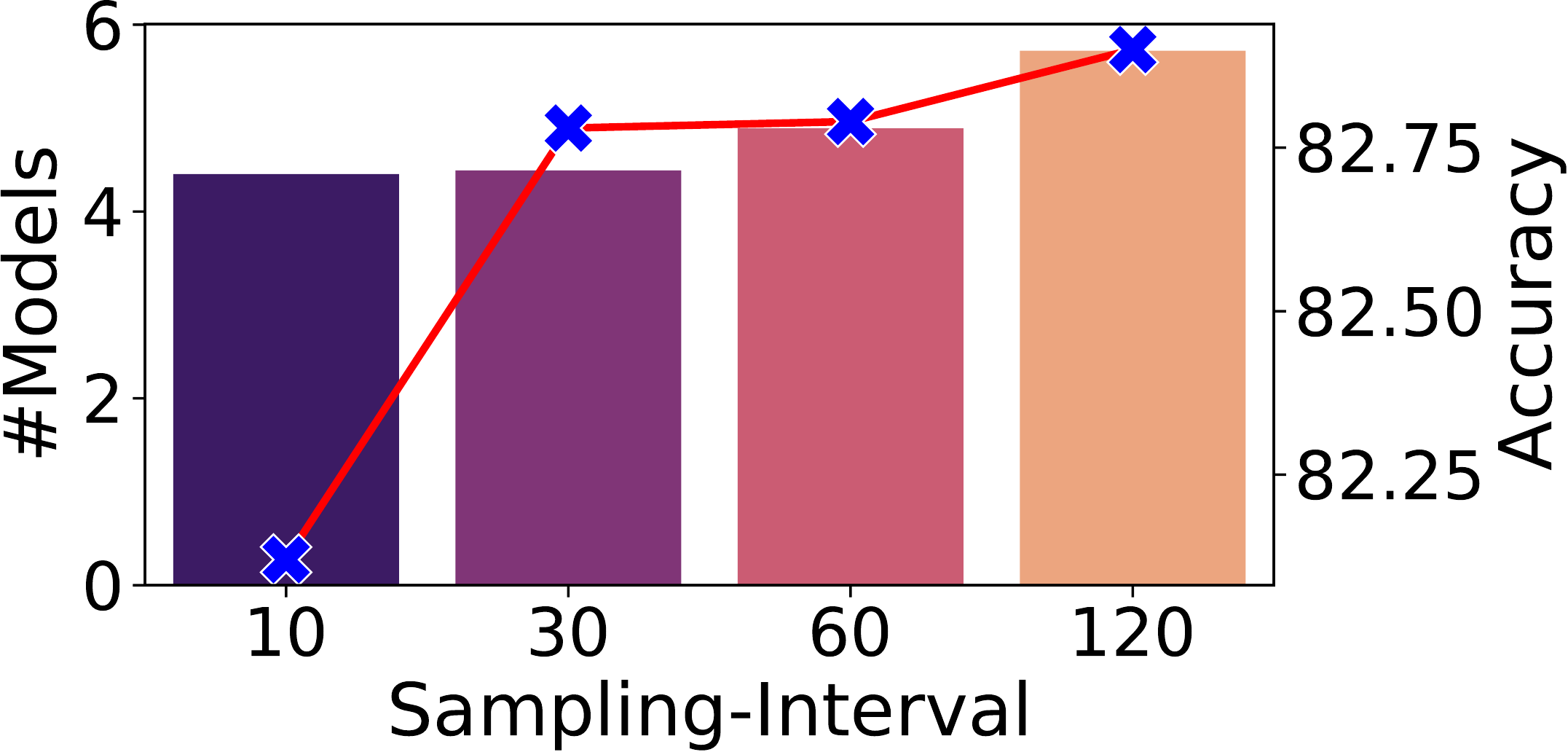}
\caption{Const-1.}
\label{fig:sens1}
\end{subfigure}
\end{minipage}
\begin{minipage}{0.9\linewidth}
\centering
\begin{subfigure}{0.99\textwidth}\includegraphics[width=0.99\textwidth]{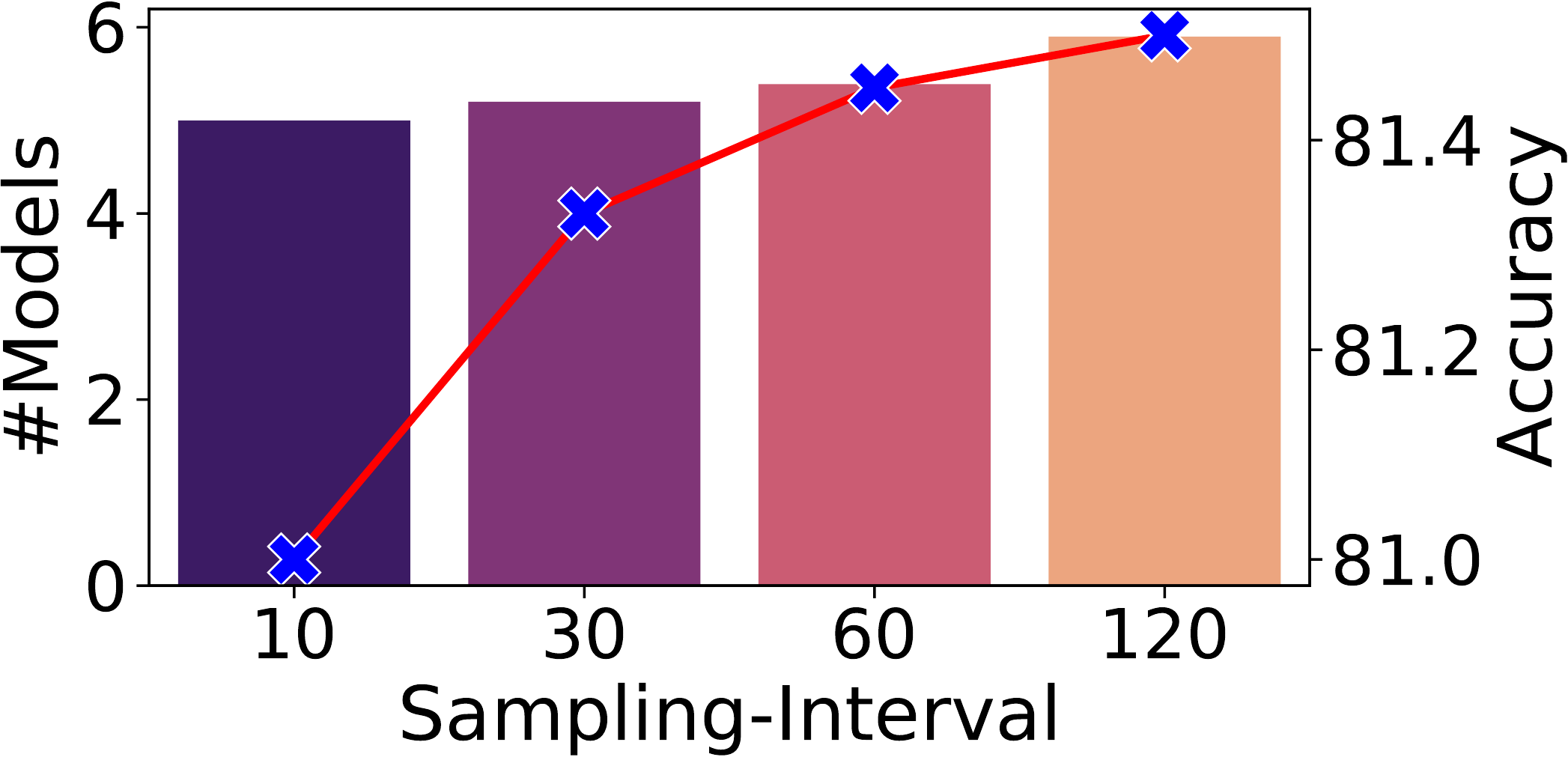}
\caption{Const-2.}
\label{fig:sens2}
\end{subfigure}
\end{minipage}
\begin{minipage}{0.9\linewidth}
\centering
\begin{subfigure}{0.99\textwidth}\includegraphics[width=0.99\textwidth]{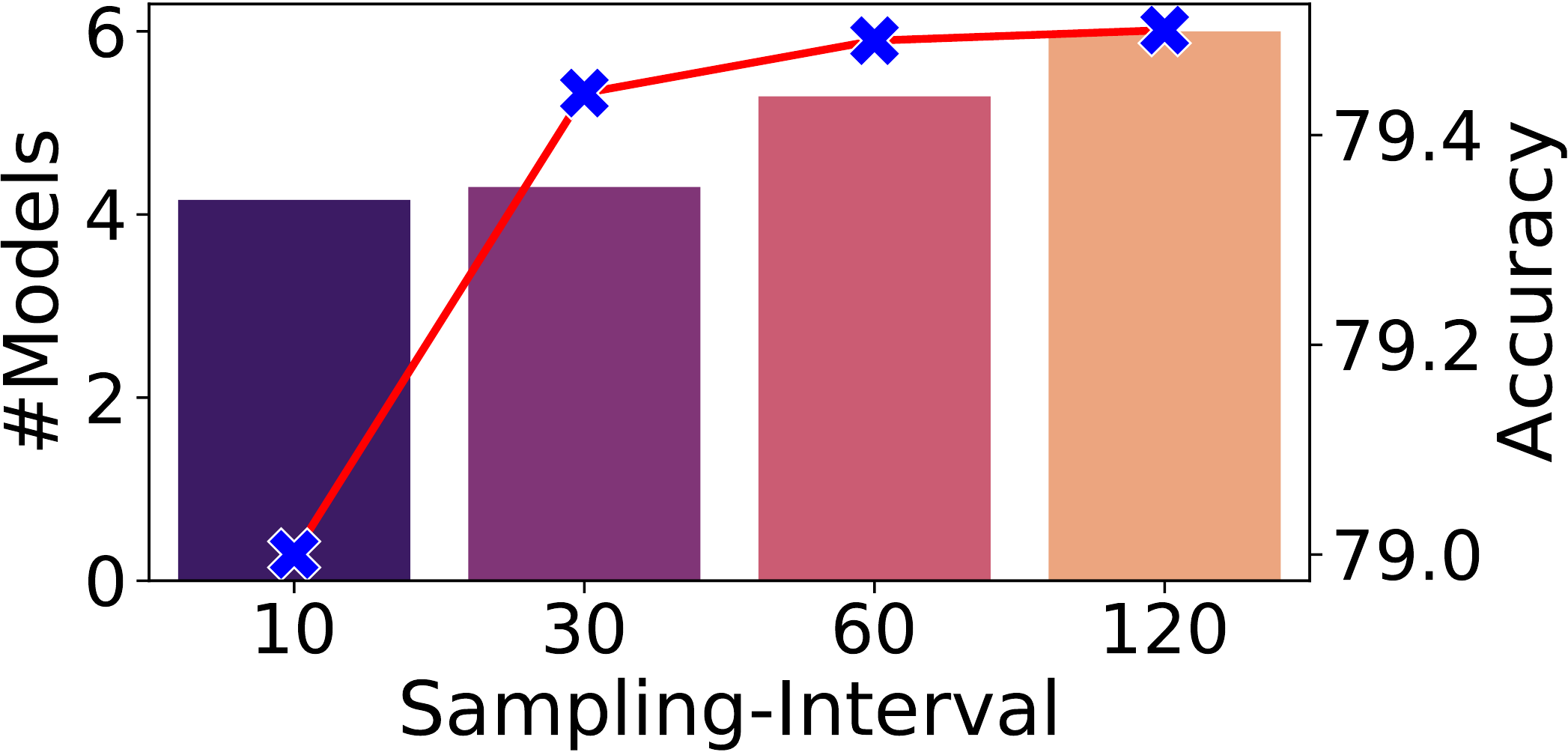}
\caption{Const-3.}
\label{fig:sens3}
\end{subfigure}
\end{minipage}
\caption{Sensitivity study.}
\label{fig:sensitivity}
\end{wrapfigure}

In this section, we analyze the sensitivity of \emph{Cocktail} with respect to various design choices including sampling interval, spot-instance failure rate and datasets used. \\
\textbf{Sampling Interval.} We use four different sampling intervals of 10s, 30s, 60s and 120s respectively. Figure~\ref{fig:sensitivity}
plots the average number of models (bar- left y-axis) and cumulative accuracy (line- right y-axis) for the different sampling intervals for queries with three different constraints. {\color{black} The 30s interval strikes the right balance with less than 0.2\% loss in accuracy and has average number models much lesser than other intervals.} Increasing the interval leads to lesser scaling operations thus results in a bigger ensemble. Thus, the 120s interval has the highest number of models.
\subsubsection{Cocktail Failure Resilience}
\begin{wrapfigure}{l}{0.25\textwidth}
\centering
\includegraphics[width=0.25\textwidth]{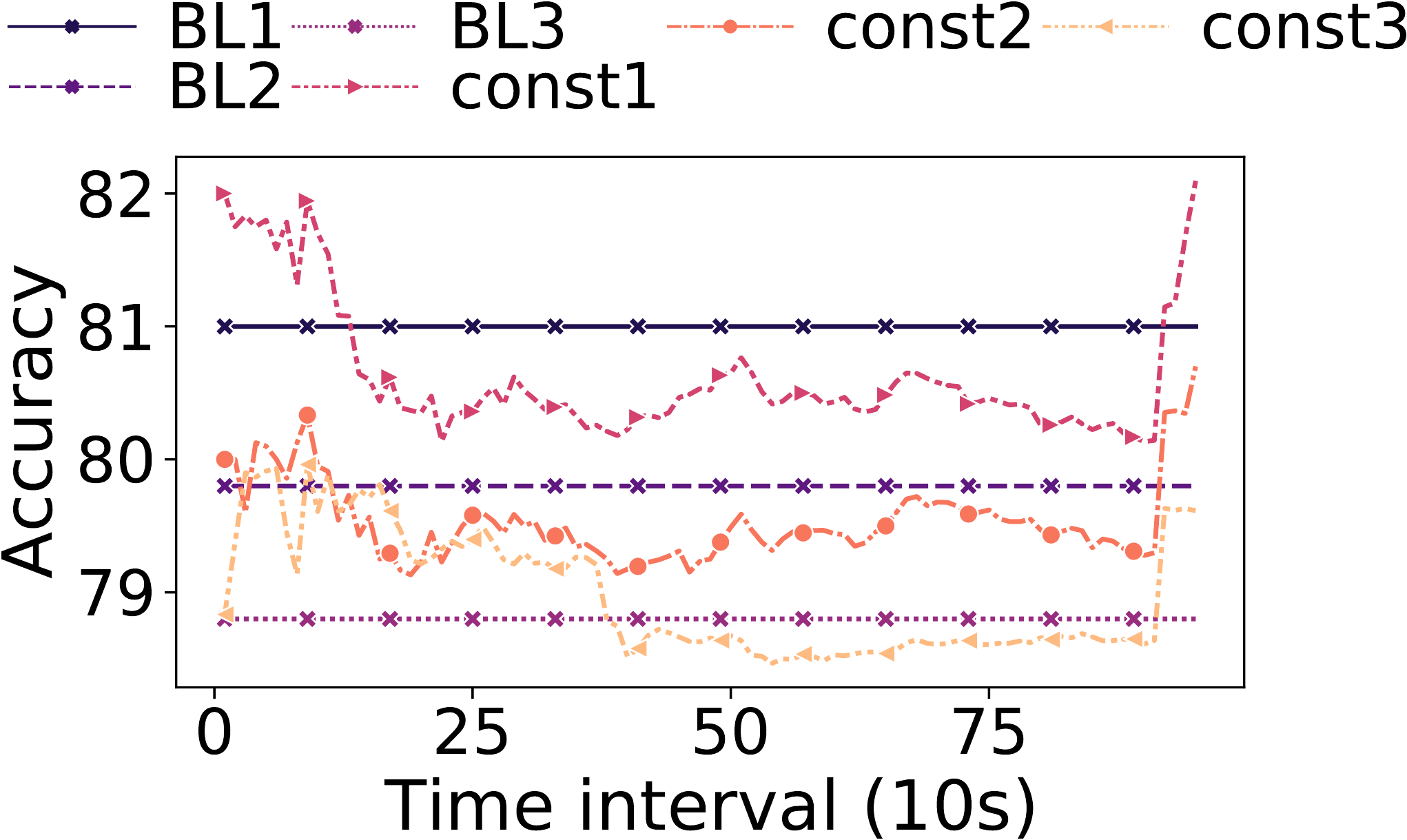}
\caption{Failure Analysis.}
\label{fig:failure}
\end{wrapfigure}

We use spot instances to host models in \emph{Cocktail}. As previously discussed in Section~\ref{sec:failure}, spot instances interruptions can lead to intermittent loss in accuracy as certain models will be unavailable in the ensemble. However for large ensembles (5 models are more), the intermittent accuracy loss is very low. Figure~\ref{fig:failure} plots the failure analysis results for top three constraints by comparing the ensemble accuracy to the target accuracy. 
The desired accuracy for all three constraints are plotted as BL1, BL2 and BL3. We induce failures in the instances using \emph{chaosmonkey}~\cite{chaosmonley} tool with a 20\% failure probability. It can be seen that queries in all three constraints suffer an intermittent loss in accuracy of 0.6\% between the time period 240\texttt{s} and 800\texttt{s}. Beyond 800\texttt{s}, they quickly recover back to the required accuracy because additional instances are spawned in place of failed instances. However, in the case of \emph{InFaas}, this would lead to 1\% failed requests due to requests being dropped from the failed instances. 

An alternate solution would be to restart the queries in running instances but that leads to increased latencies for the 1\% requests. In contrast, \emph{Cocktail} incurs a modest accuracy loss of well within 0.6\% and quickly adapts to reach the target accuracy. Thus, \emph{Cocktail} is inherently fault-tolerant owing to the parallel nature in computing multiple inferences for a single request.  We observe similar accuracy loss or lower for different probability failures of 5\%, 10\% and 25\%, respectively (results/charts omitted in the interest of space). \\
\textbf{Discussion:} For applications that are latency tolerant, we can potentially redirect requests from failed instances to existing instances, which would lead to increased tail latency.  The results we how are only for latency intolerant applications. Note that, the ensembles used in our experiments are at-least 4 models or more. For smaller ensembles, instance failures might lead to higher accuracy loss, but in our experiments, single models typically satisfy their constraints.
{\color{black}
\subsubsection{Sensitivity to Constraints}
Figure~\ref{fig:constraints-sensitivity} plots the sensitivity of model selection policy under a wide-range of latency and accuracy constraints. In Figure~\ref{fig:fixed-accuracy}, we vary the latency under six different constant accuracy categories. It can be seen that for fixed accuracy of 72\%, 78\% and 80\%,  the average number of models increase with increase in latency, but drops to 1 for the highest latency.  Intuitively, singe large models with higher latency can satisfy the accuracy, while short latency models need to be ensembled to reach the same accuracy. For accuracy greater than 80\%, the ensemble size drops with higher latencies. This is because the models which offer higher accuracy are typically dense and hence, smaller ensembles are sufficient. In Figure~\ref{fig:fixed-latency}, we vary the accuracy under six different constant latency categories. It can be seen that for higher accuracies, \emph{Cocktail} tries to ensemble more models to reach the accuracy, while for lower accuracy it resorts to using single models.
\begin{figure}
\begin{minipage}[t]{0.99\linewidth}
\begin{subfigure}[t]{.47\textwidth}
\centering
\includegraphics[width=0.99\textwidth]{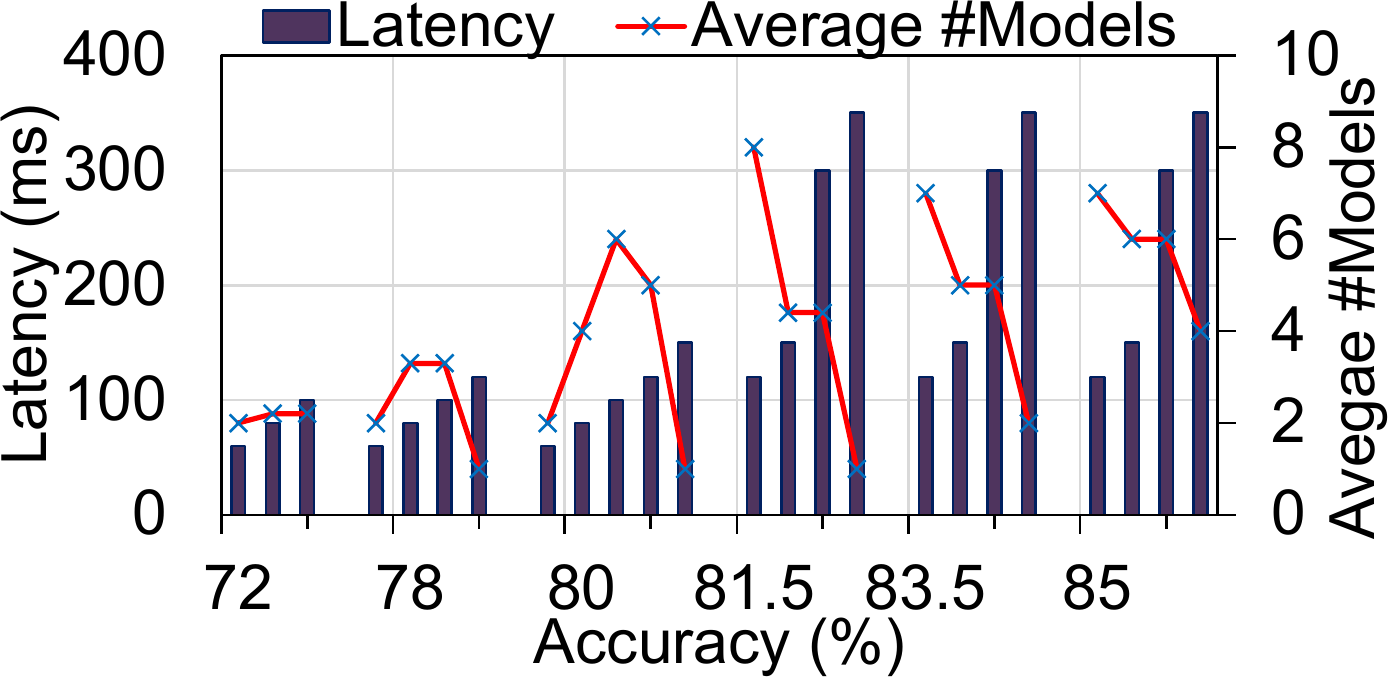}
\caption{Fixed Accuracy.}
\label{fig:fixed-accuracy}
\end{subfigure}
\begin{subfigure}[t]{0.47\textwidth}
\centering
\includegraphics[width=0.99\textwidth]{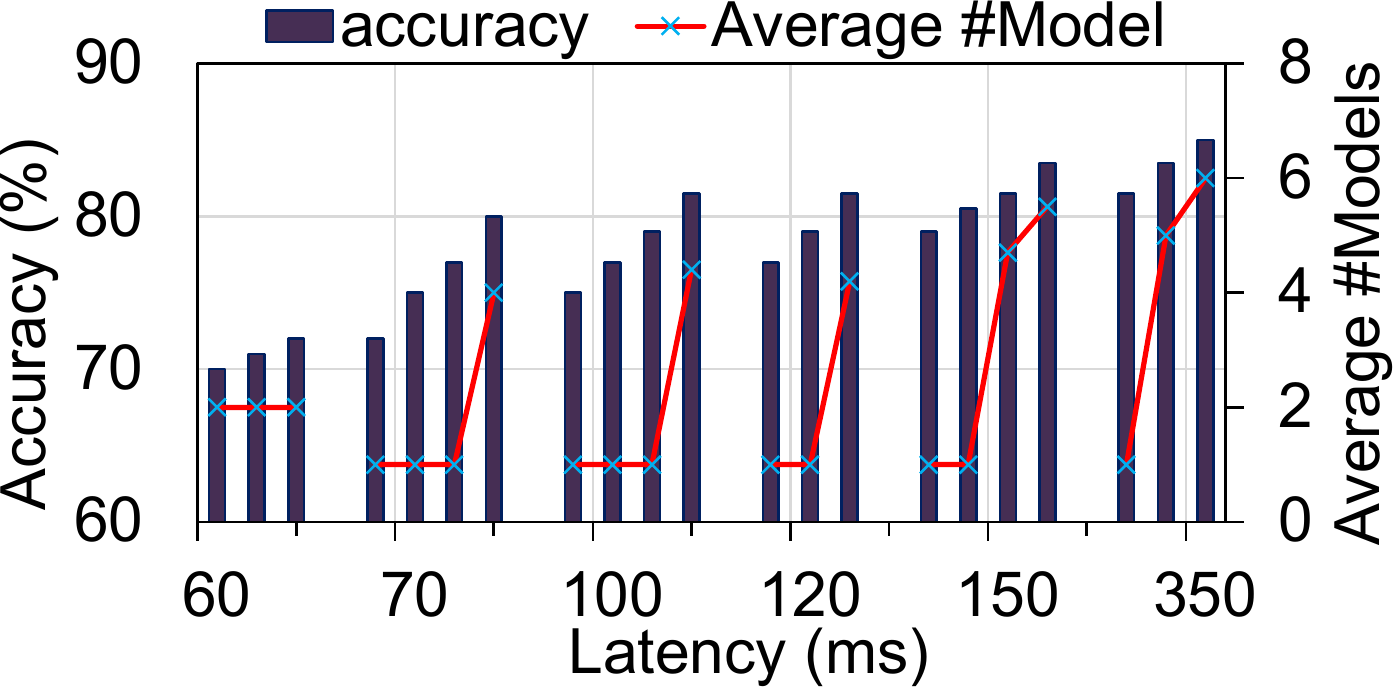}
\caption{Fixed Latency.}
\label{fig:fixed-latency}
\end{subfigure}
\end{minipage}
\caption{{\color{black}Sensitivity Constraints under fixed latency and accuracy. Bar graphs (latency) plotted using primary y-axis and line graph (\#models) plotted using secondary y-axis.}}
\label{fig:constraints-sensitivity}
\end{figure}

\begin{figure}[t]
\begin{minipage}{0.99\linewidth}
\begin{subfigure}{.48\textwidth}
\centering
\includegraphics[width=0.99\textwidth]{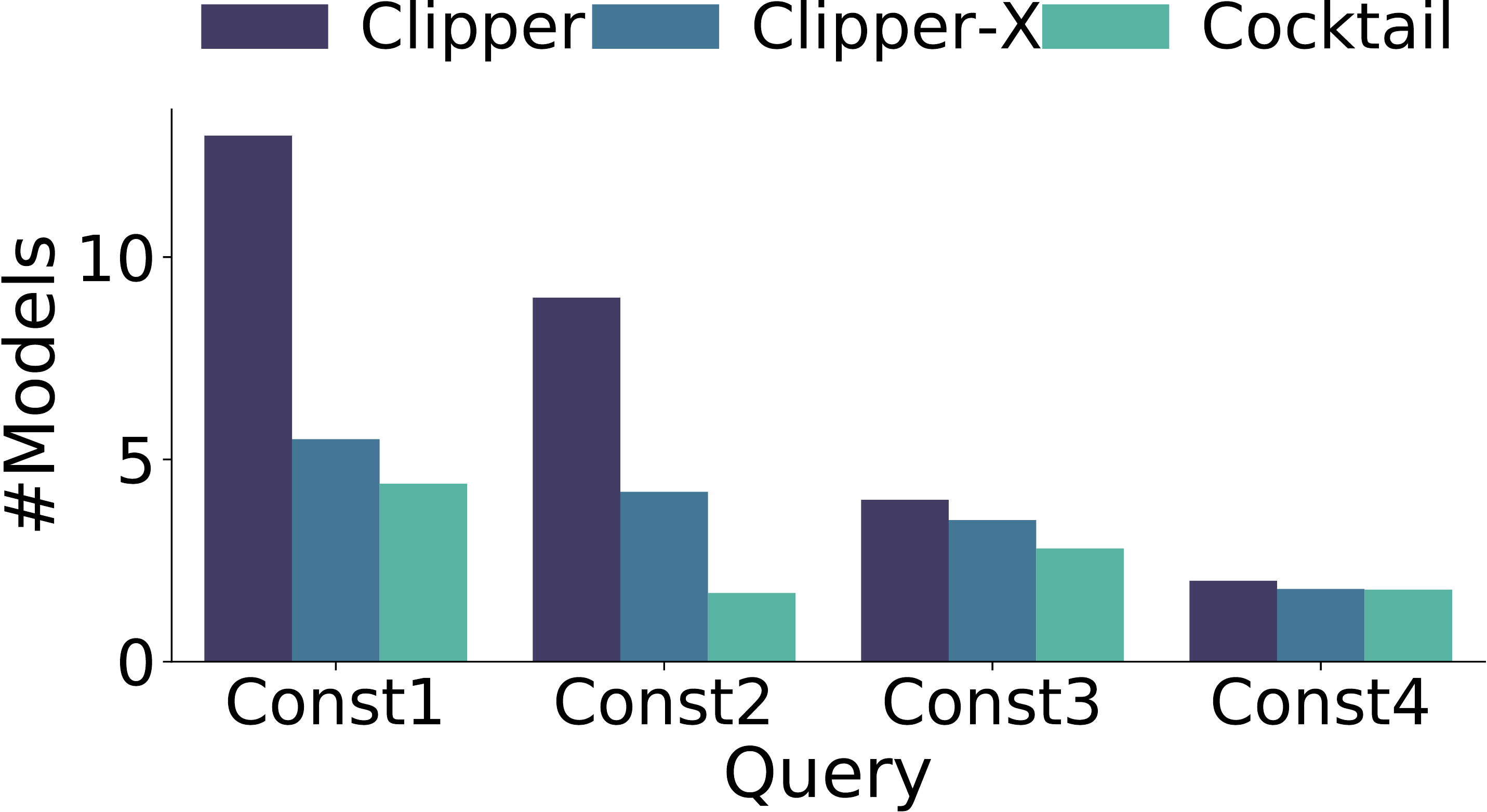}
\caption{Image Classification-Cifar-100.}
\label{fig:cifar-models}
\end{subfigure}
\begin{subfigure}{0.48\textwidth}
\centering
 \includegraphics[width=0.99\textwidth]{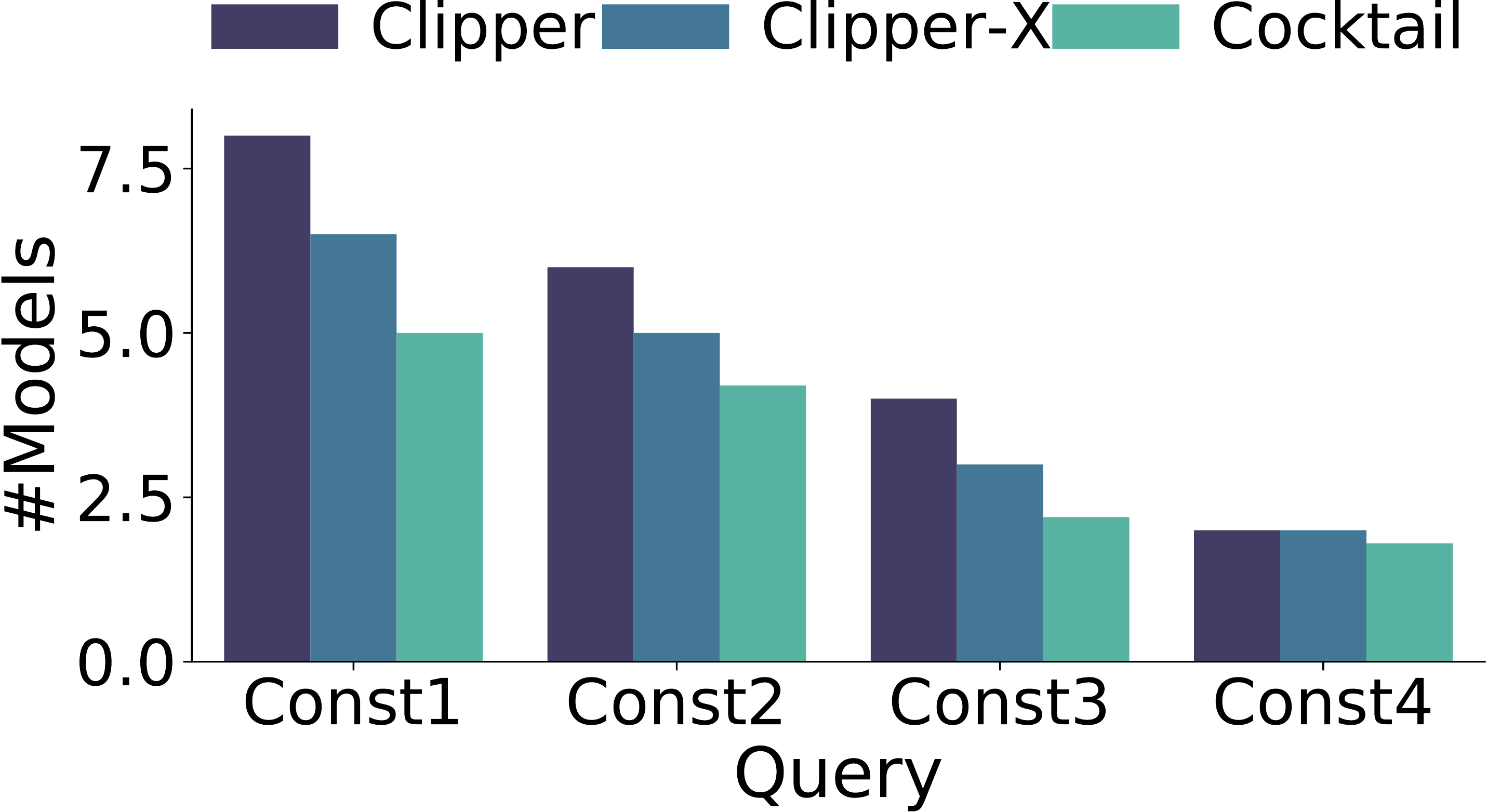}
\caption{Sentiment analysis.}
\label{fig:bert-models}
\end{subfigure}
\end{minipage}
\caption{{\color{black}Average number of models used in the ensemble.}}
\label{fig:models-cifar}
\end{figure}

\begin{figure}[t]
\begin{minipage}{0.99\linewidth}
\begin{subfigure}{0.48\textwidth}
\centering
 \includegraphics[width=0.99\textwidth]{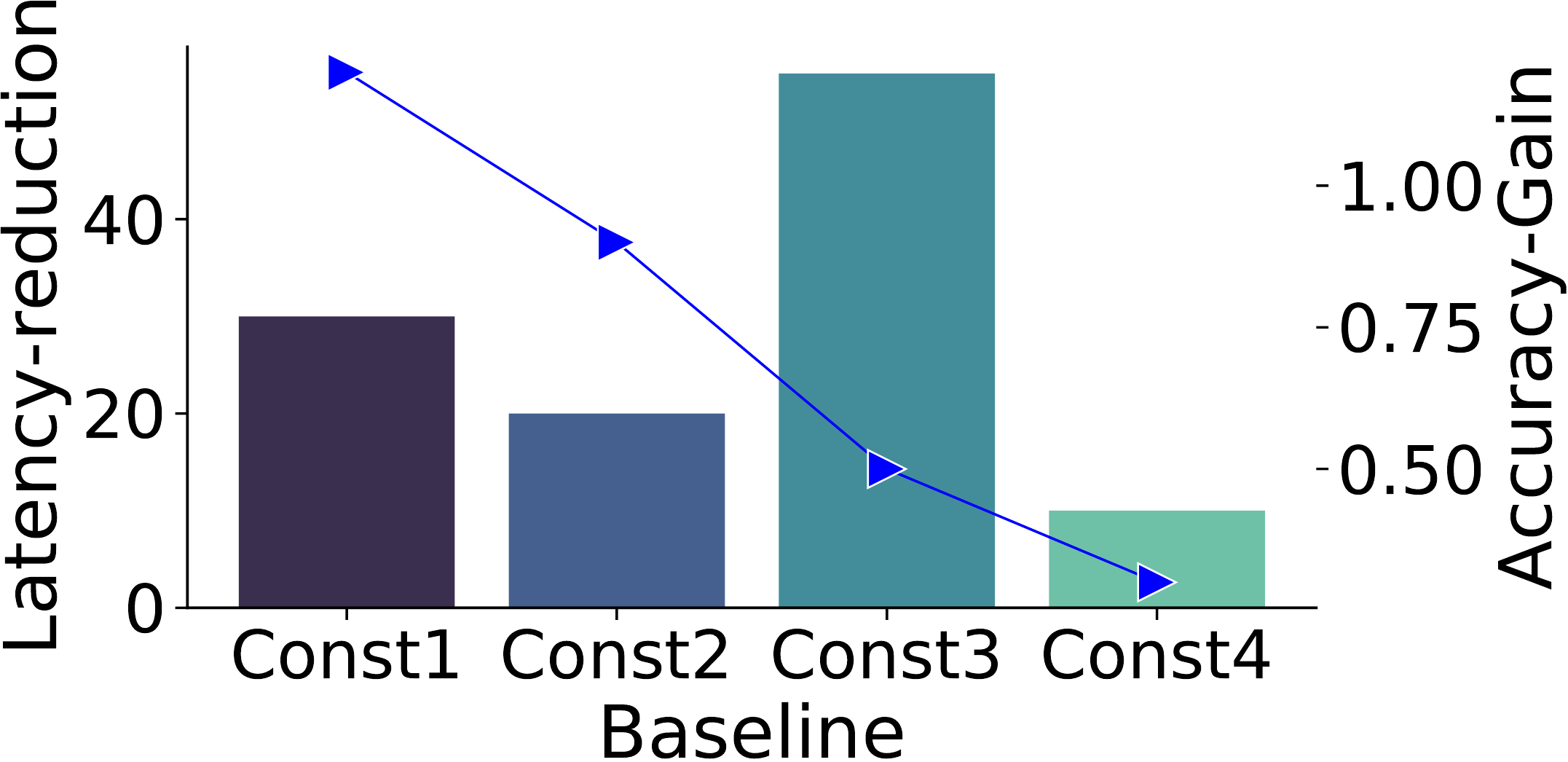}
\caption{Image Classification:Cifar100.}
\label{fig:latency-cifar}
\end{subfigure}
\begin{subfigure}{.48\textwidth}
\centering
\includegraphics[width=0.99\textwidth]{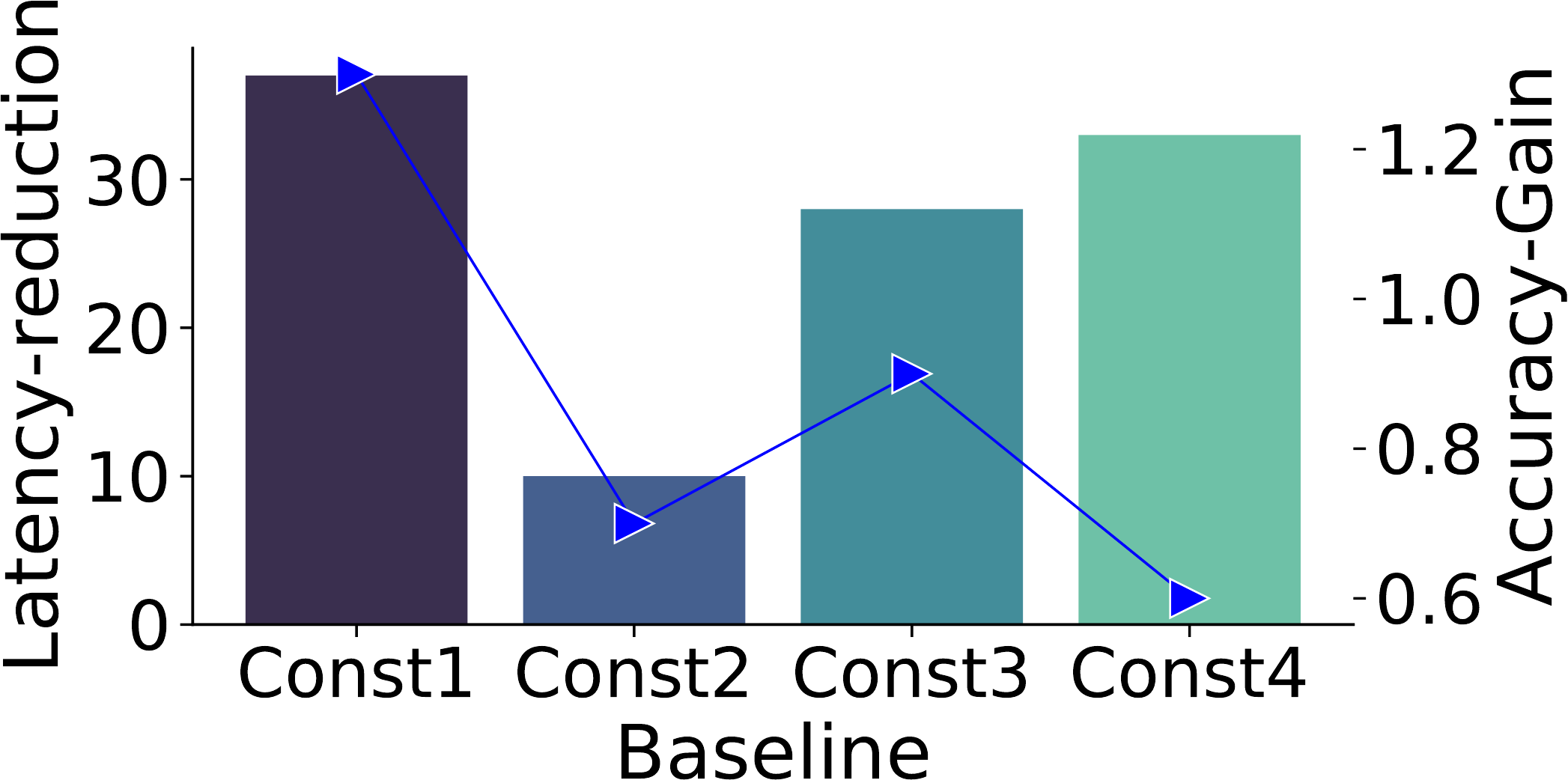}
\caption{Sentiment Analysis.}
\label{fig:latency-bert}
\end{subfigure}
\end{minipage}
\caption{{\color{black}Latency reduction (\%) plotted as bar graph(primary y-axis) and accuracy gains (\%) plotted as line graph (secondary y-axis) over InFaaS.}}
\label{fig:latency-acc}
\end{figure}
\begin{figure}[ht]
\begin{minipage}{0.99\linewidth}
\begin{subfigure}{.48\textwidth}
\centering
\includegraphics[width=0.99\textwidth]{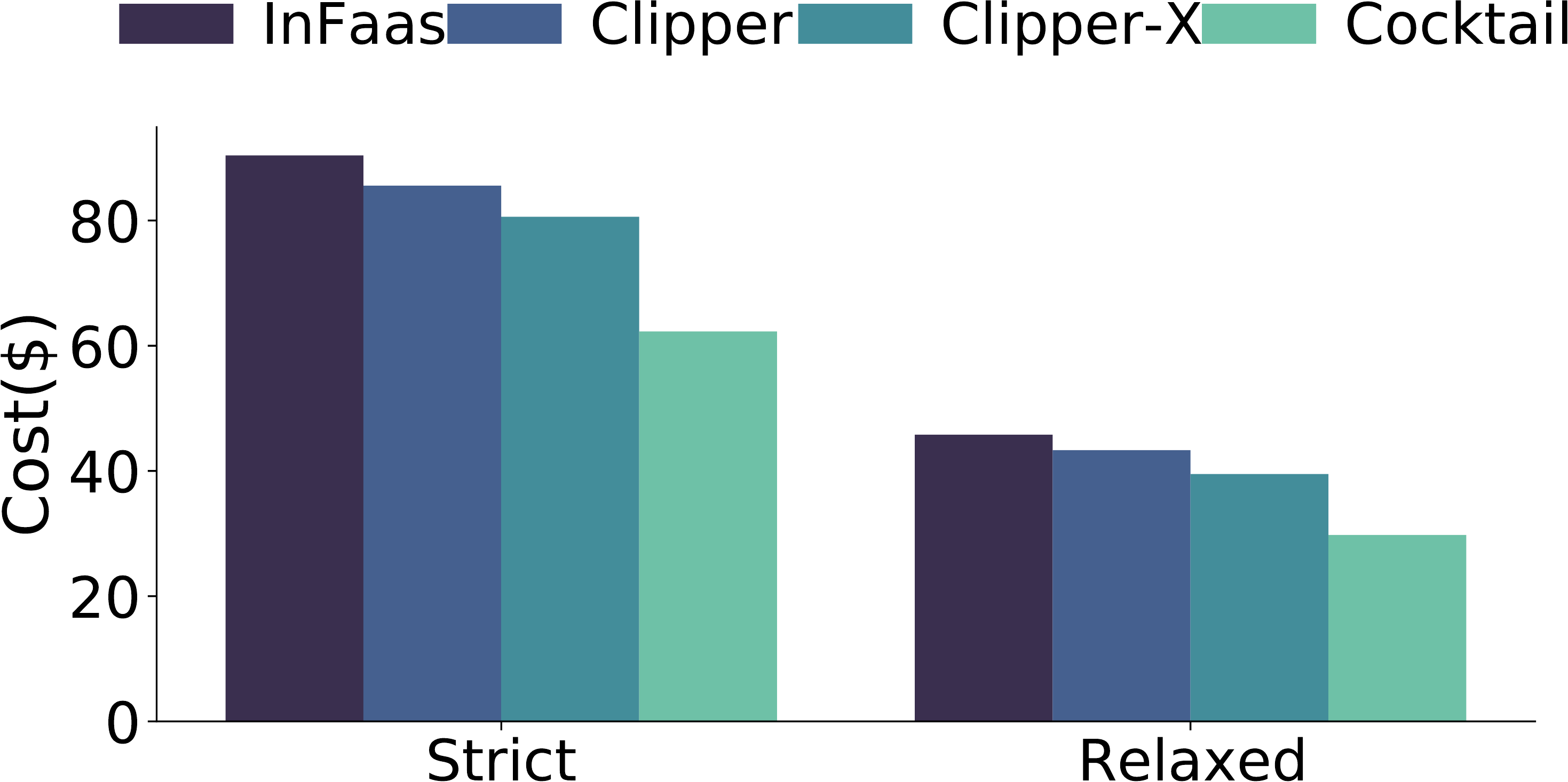}
\caption{Wiki Trace.}
\label{fig:wiki-cost-bert}
\end{subfigure}
\begin{subfigure}{0.48\textwidth}
\centering
 \includegraphics[width=0.99\textwidth]{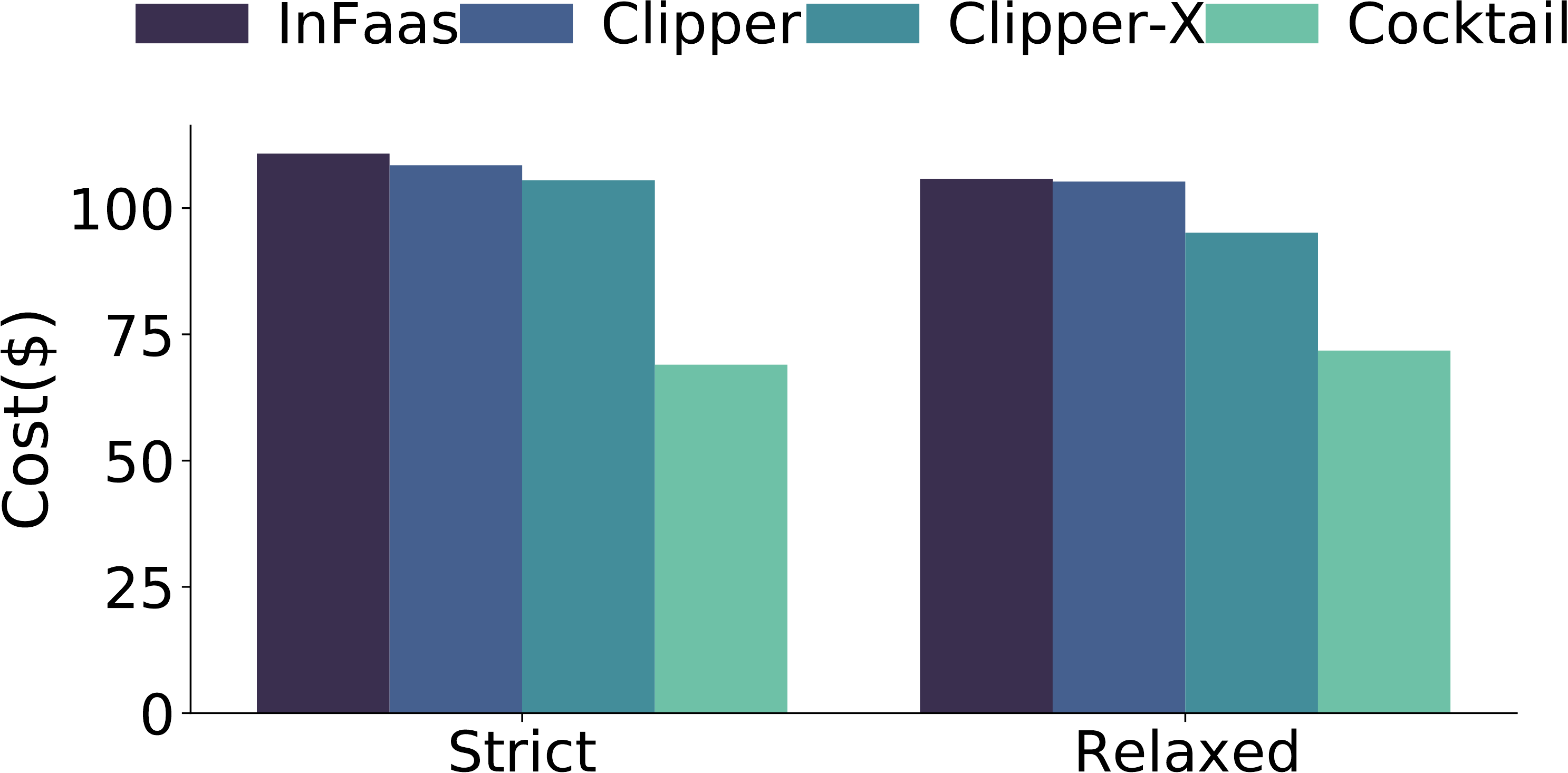}
\caption{Twitter Trace.}
\label{fig:twitter-cost-bert}
\end{subfigure}
\end{minipage}
\caption{{\color{black}Cost savings of Cocktail for Sentiment Analysis.}}
\label{fig:cost-bert}
\end{figure}
\subsubsection{Sensitivity to Dataset}
To demonstrate the applicability of \emph{Cocktail} to multiple datasets, we conducted similar experiments as elucidated in Section~\ref{sec:eval} using the \texttt{CIFAR-100} dataset~\cite{cifar}. It comprises of 100 distinct image classes and  we trained 11 different models including the nine that are common from Table~\ref{tbl:models}. Figure~\ref{fig:cifar-models} plots the average number of models used by the three policies for the top four constraints. It can be seen that \emph{Cocktail} shows similar reduction (as Imagenet) while using only 4.4 models on average. As expected, \emph{Clipper} and \emph{Clipper-X} use more models than \emph{Cocktail} (11 and 5.4, respectively) due to non-aggressive scaling down of the models used. 

Figure~\ref{fig:latency-cifar} plots the latency reduction and accuracy boost when compared to \emph{InFaaS} (baseline). While able to reduce 60\% of the models used in the ensemble, \emph{Cocktail} also reduces latency by up to 50\% and boosts accuracy by up to 1.2\%. \emph{Cocktail} was also able to deliver modest accuracy gain of 0.5\% than \emph{Clipper} (not plotted). The accuracy gain seen in \texttt{CIFAR-100} is lesser than ImageNet dataset because the class-based weighted voting works effectively when handling large number of classes (100 in \texttt{CIFAR} vs 1000 in ImageNet). Nevertheless, \emph{Cocktail} is able to deliver the accuracy at 2x lower latency than \emph{InFaaS} and 1.35x lower cost than Clipper.

\subsection{General Applicability of Cocktail}

To demonstrate the general applicability of \emph{Cocktail} to other classification tasks, we evaluated \emph{Cocktail} using a Sentiment Analysis application for two datasets. The results reported are averaged across both the datasets. Figure~\ref{fig:bert-models} plots the average number of models used by the three policies for the top four constraints. As shown for Const-1, \emph{Cocktail} shows similar reduction (as image-classification) with only using 4.8 models on average, which is 40\% and 26\% lower than \emph{Clipper} and \emph{Clipper-X},  respectively. \emph{Cocktail} is also able to reduce the number of models by 30\% and 50\% for medium ensembles (Const-2 \& Const-3) as well. 

Figure~\ref{fig:latency-bert} plots the latency reduction and accuracy gain, compared to \emph{InFaaS} (baseline). While being able to reduce 50\% of the models used in the ensemble, \emph{Cocktail} also reduces latency by up to 50\% and improves accuracy by up to 1.3\%. Both \emph{Cocktail} and \emph{Clipper} deliver the same overall accuracy (96\%, 94.5\%, 93.5\%, and 92\%)). Since sentiment analysis only has 2-3 classes, there are no additional accuracy gains by using the class-based weighted voting. However, the model selection policy effectively switches between different models based on the structure of input text (equivalent to classes in images). For instance, complex sentences are more accurately classified by denser models compared to smaller. Despite the lower accuracy gains, \emph{Cocktail} is able to reduce the cost (Figure~\ref{fig:cost-bert}) of model-serving by 1.45$\times$ and 1.37$\times$ for Wiki trace compared to \emph{InFaaS} and \emph{Clipper}, respectively. 
}



\section{Concluding Remarks}
\label{sec:conclusion}
There is an imminent need to develop model serving systems that can deliver highly accurate, low latency predictions at reduced cost. In this paper, we propose and evaluate \emph{Cocktail}, a cost-effective model serving system that exploits ensembling techniques to meet high accuracy under low latency goals. In \emph{Cocktail}, we adopt a three-fold approach to reduce the resource footprint of model ensembling. More specifically, we (i) develop a novel dynamic model selection, (ii) design a prudent resource management scheme that utilizes weighted autoscaling for efficient resource allocation, and (iii)  leverage transient VM instances to reduce the deployment costs. Our results from extensive evaluations using both CPU and GPU instances on AWS EC2 cloud platform demonstrate that \emph{Cocktail} can reduce deployment cost by 1.4$\times$, while reducing latency by 2$\times$ and satisfying accuracy for 96\% of requests, compared to state-of-the-art model-serving systems. 

\bibliographystyle{plain}
{\footnotesize
\bibliography{references}}
\newpage
\appendix
\section*{Appendix}
\label{sec:appendix}
\section{Modeling of Ensembling}
While performing an ensemble it is important to be sure that we can reach the desired accuracy by combining more models. In our design, we solve our first objective function (described in Section~\ref{sub:scheme:DynModel}) by combining all available models which meet the latency SLO. To be sure that the combination will give us the desired accuracy of the larger model, we try to theoretically analyse the scenario. We formulate the problem conservatively as following. 

{We perform an inference by ensembling 'N' models, and each of these models have accuracy 'a'. Therefore the probability of any model giving a correct classification is 'a'. We assume the output to be correct if majority of them, i.e. $\lfloor N/2 \rfloor +1$ of them give the same result. Then, the final accuracy of this ensemble would be the probability of at least $\lfloor N/2 \rfloor +1$ of them giving a correct result.} 

To we model this problem as a coin-toss problem involving $N$ biased coins with having probability of occurrence of head to be $a$. Relating this to our problem, each coin represents a model, and an occurrence of head represents the model giving the correct classification. Hence, the problem boils down to find the probability of at least $\lfloor N/2 \rfloor +1$ heads when all N coins are tossed together. This is a standard binomial distribution problem and can be solved by using the following formula: 
\begin{align*}
P_{head}  = \sum_{i=\lfloor \frac{N}{2}\rfloor +1}^N {\binom{N}{i}} \hspace{1mm }a^{i} \hspace{1mm}(1-a)^{(N-i)}.
\end{align*}

To further quantify, let us consider the case where we need to determine if we can reach the accuracy of NasNetLarge (82\%) by combining rest of the smaller models which have lesser latency than NasNetLarge. We have 10 (therefore N = 10) such models and among them the least accurate model is MobileNetV1 (accuracy 70\%, therefore a = 0.70). We need to find the probability of at least 6 of them being correct. Using the equation above we find the probability to be 
\begin{align*}
P_{head}  = \sum_{i=\lfloor \frac{10}{2} \rfloor +1 = 6}^{10} {\binom{10}{i}} \hspace{1mm }0.7^{i} \hspace{1mm}(1-0.7)^{(10-i)} = 0.83 
\end{align*}
This corresponds to an accuracy of 83\%, which is greater than our required accuracy of 82\%). Given all the other models have higher accuracy, the least accuracy we can expect with such an ensemble is 83\%. This analysis forms the base of our ensemble technique, and hence proving the combination of multiple available models can be more accurate than the most accurate individual model. 
\section{Why DeepARest Model?}
We quantitatively justify the choice of using DeepARest by conducting a brick-by-brick comparison of the accuracy loss when compared with other state-of-the-art prediction models used in prior work. 
\begin{table}[h]
\footnotesize
\centering
\begin{tabular}{l|l}
\hline
\textbf{Model} & \textbf{RMSE} \\ \hline
MWA & 77.5 \\ \hline
EWMA & 88.25 \\ \hline
Linear R. & 87.5 \\ \hline
Logsitic R. & 78.34 \\ \hline
Simple FF. & 45.45 \\ \hline
DeepArEst & 26.67 \\ \hline
LSTM & 28.56 \\ \hline
\end{tabular}
\caption{}
\label{tbl:predictions}
\end{table}
Table~\ref{tbl:predictions} shows the root mean squared error (RMSE) incurred by all the models. The ML models used in these experiments are pre-trained with 60\% of the Twitter arrival trace. It is evident that the LSTM and DeepAREst have lowest RMSE value. DeepARest is 10\% better than LSTM model. Since the primary contribution in \emph{Cocktail} is to provide high accuracy and low latency predictions at cheaper cost, application developers can adapt the prediction algorithm to their needs or even plug-in their own prediction models.

\section{System Overheads}
We characterize the system-level overheads incurred due to the design choices in \emph{Cocktail}. The \emph{mongodb} database is a centralized server, which resides on the head-node. We measure the overall average latency incurred due to all reads/writes in the database, which is well within 1.5ms. The DeepARest prediction model which is not in the critical decision-making path runs as a background process incurring 2.2 ms latency on average. The weighted majority voting takes 0.5ms and the model selection policy takes 0.7ms. The time taken to spawn new VM takes about 60s to 100s depending on the size of the VM instance. The time taken to choose models from the model-cache is less than 1ms. The end-to-end response time to send the image to a worker VM and get the prediction back, was dominated by about 300ms (at maximum) of payload transfer time.

\section{Instance configuration and Pricing}
\begin{table}[hbtp]
\footnotesize
\centering
\begin{tabular}{|l|c|c|c|}
\hline
\textbf{Instance} & \multicolumn{1}{l|}{\textbf{vCPUs}} & \multicolumn{1}{l|}{\textbf{Memory}} & \multicolumn{1}{l|}{\textbf{Price}} \\ \hline
{ C5a.xlarge} & { 4} & { 8 GiB} & { \$0.154} \\ \hline
{ C5a.2xlarge} & { 8} & { 16 GiB} & { \$0.308} \\ \hline
{ C5a.4xlarge} & { 16} & { 32 GiB} & { \$0.616} \\ \hline
{ C5a.8xlarge} & { 32} & { 64 GiB} & { \$1.232} \\ \hline
\end{tabular}
\caption{Configuration and Pricing for EC2 C5 instances.}
\label{tab:instances}
\end{table}

\section{Sentiment Analysis Models}
\begin{table}[h]
\centering
\footnotesize
\resizebox{0.45\textwidth}{!}{%
\begin{tabular}{||l|c|c|c|l||}
\hline
\textbf{Model} & \textbf{\begin{tabular}[c]{@{}c@{}}Params \\ (M)\end{tabular}} & \textbf{\begin{tabular}[c]{@{}c@{}}Top-1\\ Accuracy(\%)\end{tabular}} & \textbf{\begin{tabular}[c]{@{}c@{}}Latency \\ (ms)\end{tabular}} & \textbf{\begin{tabular}[c]{@{}l@{}}$P_f$\end{tabular}} \\ \hline
Albert-base~\cite{albert} & 11 & 91.4 & 55 & 7 \\ \hline
CodeBert~\cite{codebert} & {\color[HTML]{263238} 125} & {\color[HTML]{263238} 89} & 79 & 6 \\ \hline
DistilBert~\cite{distilbert} & 66 & {\color[HTML]{263238} 90.6} & 92 & 5 \\ \hline
Albert-large & {\color[HTML]{263238} 17} & {\color[HTML]{263238} 92.5} & 120 & 4 \\ \hline
XLNet~\cite{xlnet} & {\color[HTML]{263238} 110} & {\color[HTML]{263238} 94.6} & 165 & 3 \\ \hline
Bert~\cite{bert} & 110 & {\color[HTML]{263238} 92} & 185 & 3 \\ \hline
Roberta~\cite{roberta}  & {\color[HTML]{263238} 355} & {\color[HTML]{263238} 94.3} & 200 & 2 \\ \hline
Albert-xlarge & {\color[HTML]{263238} 58} & {\color[HTML]{263238} 93.8} & 220 & 1 \\ \hline
Albert-xxlarge & {\color[HTML]{263238} 223} & {\color[HTML]{263238} 95.9} & 350 & 1 \\ \hline
\end{tabular}
}
\caption{Pretrained models for Sentiment Analysis using BERT.}
\label{tbl:bert-models}
\end{table}
\section{Spot Instance Price Variation}
We profile the spot price of 4 types of \texttt{C5} EC2 VMs over a 2-week period in August 2020.
\begin{figure}[h]
\centering
 \includegraphics[width=0.7\linewidth]{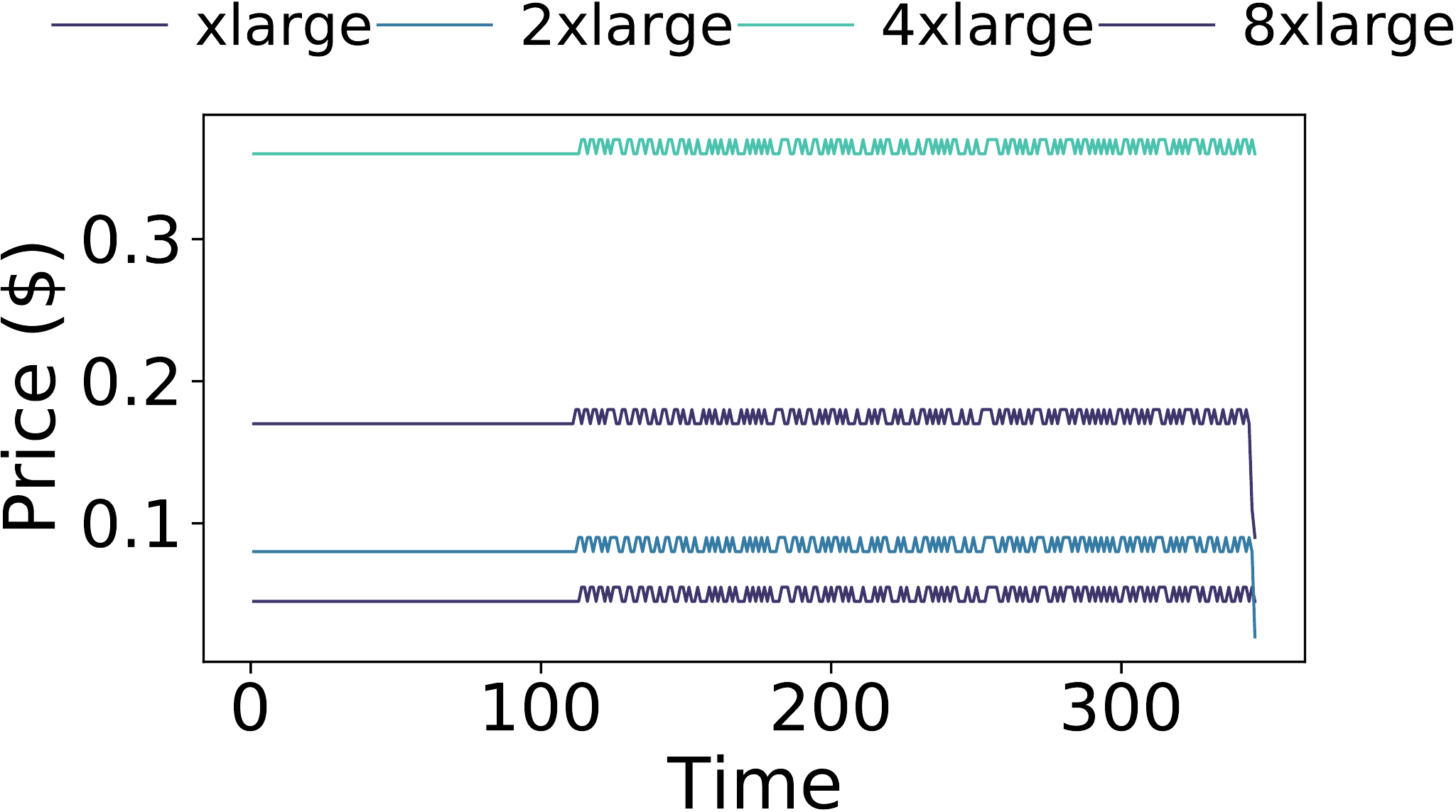}
\caption{Spot instance price variation (time is in hours).}
\label{fig:spot}
\end{figure}


\end{document}